\documentclass[preprint]{elsarticle}
\biboptions{numbers,sort&compress}
\usepackage{graphicx}
\usepackage[utf8]{inputenc}
\usepackage{blindtext}
\usepackage[scr=rsfs]{mathalpha}
\usepackage{pgfplots}
\usepackage[T1]{fontenc}
\usepackage[margin=1.0in]{geometry}
\usepackage{float}
\usepackage{fullpage}
\usepackage{amsmath}
\usepackage{array}
\usepackage{hyperref}
\usepackage{physics}
\usepackage{amsfonts}
\usepackage{lineno}
\modulolinenumbers[10]
\numberwithin{equation}{section}
\usepackage{MnSymbol}
\pgfplotsset{compat=1.15}
\usepackage{subfigure} 
\makeatletter
\def\ps@pprintTitle{%
  \let\@oddhead\@empty
  \let\@evenhead\@empty
  \let\@oddfoot\@empty
  \let\@evenfoot\@oddfoot
}
\newcommand{\tn}[1]{\ensuremath{\underline{\mathbf{#1}}}}
\newcommand{\pfrac}[2]{\ensuremath{\dfrac{\partial #1}{\partial #2}}}
\newcommand{\inner}[2]{\big< #1 ,  #2 \big>}
\newcommand{\bm}[1]{\ensuremath{\mathbf{#1}}}
\newcommand{\bs}[1]{\ensuremath{\boldsymbol{#1}}}

\makeatother
\begin{document}
\title{Scalable \emph{In Situ} Compression of Transient Simulation Data Using Time-Dependent Bases }
\begin{frontmatter}
\author[1]{Shaghayegh Zamani Ashtiani}
%\address[1]{Department of Mechanical Engineering, University of Pittsburgh, Pittsburgh, PA-15206.}
\author[2]{Mujeeb R.Malik}
\author[1]{Hessam Babaee\corref{cor1}}
\address[1]{Department of Mechanical Engineering, University of Pittsburgh, Pittsburgh, PA-15206.}
\address[2]{NASA Langley Research Center, Hampton, VA-23681}
\ead{h.babaee@pitt.edu}

\begin{abstract}
Large-scale simulations  of time-dependent problems generate a massive amount of data and with the explosive increase in computational resources  the size of the data generated by these simulations has increased significantly. This has imposed severe limitations on the amount of data that can be stored and has elevated the issue of input/output (I/O) into one of the major bottlenecks of high-performance computing.    In this work, we present an \emph{in situ} compression technique to reduce the size of the data storage by orders of magnitude. This methodology is based on time-dependent subspaces and it  extracts low-rank structures from multidimensional streaming data by decomposing the data  into a set of   time-dependent bases and a core tensor. We derive closed-form evolution equations for the core tensor as well as the time-dependent bases. The presented methodology does not require the data history  and the computational cost of its extractions scales linearly with the size of data --- making it suitable for large scale streaming datasets.  To control the compression error, we present an  adaptive strategy to add/remove modes to maintain the reconstruction error below a given threshold.  We present four demonstration cases: (i) analytical example, (ii)  incompressible unsteady reactive flow, (iii) stochastic turbulent reactive flow, and (iv) three-dimensional turbulent channel flow. 
\end{abstract}
\begin{keyword}
Tensor decomposition \sep Real-time computation \sep Time-dependent bases \sep \emph{in situ} compression
\end{keyword}
\end{frontmatter}

\section{Introduction}\label{sec:Intro}
The ability to perform large-scale simulations has grown explosively in the past few decades and it will only continue to grow in the near future. On the other hand, our ability to effectively analyze or in some applications even store the  data that these simulations generate is lagging behind --- impeding many scientific discoveries \cite{NASA2030vision,NASA2030progress,Appliedmatreport2014}. 
% Extraction of low-dimensional structure from  high dimensional models/data is a powerful tool for analysis, prediction and control of vast array of applications. In the context of scientific applications, the low-dimensional structures can be used  to: (i)  compress the data; (ii) discover intrinsic variables and thus enabling a parsimonious and interpretable description of complex high-dimensional systems, and (iii)  build  reduced order models (ROMs) to reduce the cost of uncertainty propagation, optimization and solving inverse problems.  
For extreme-scale simulations the sheer size of the data  imposes  input/output (I/O) constraints that impede  storing temporally-resolved simulation data \cite{DOE_21}.  An example is the direct numerical simulation (DNS) of turbulent combustion, in which the creation of an ignition kernel -- an intermittent phenomenon -- occurs on the order of 10 simulation time steps, while typically every 400th time step is stored to maintain the I/O overhead at a reasonable level \cite{BAB12}.  Similarly, climate simulations use time steps of minutes but model outputs are typically written at daily to monthly intervals, limiting the ability to understand extreme weather events and projections of their future changes.  The I/O and network limitations motivate for a paradigm shift from \emph{postprocess-centric} to \emph{concurrent} data analysis, wherein  the raw  data is analysed as it is generated in an \emph{in situ} or \emph{in-transit} framework and  given the size of the  data and the demand for real-time analysis, only scalable  algorithms with low computational complexity have a chance of being feasible. \\

Being able to store time-dependent simulation data with higher temporal resolution is critically important for many purposes. One of the most common applications is checkpointing data for simulation restarts, where the simulation may need to restart from the last snapshot \cite{checkpoint}. This is particularly important  for   parallel simulations with a large number of  computational nodes with long runtimes,  where node failure can interrupt the simulation or queue limits may be exceeded.  Frequent checkpointing lowers the cost of loss in computational resources. The other applications are for data analysis, including visualization and extracting coherent structures from the simulation.\\    

 Data compression techniques can be broadly divided to \emph{lossy} and \emph{lossless} classes \cite{li2018data}. In the lossless compression, the compressed data has no error when compared to the decompressed data. Examples of lossless compression techniques are FPZIP \cite{lindstrom2006fast} and  ACE \cite{fout2012adaptive}. All of these techniques require a significant amount of memory and they are not suitable for large-scale simulations  where only on-the-fly compression techniques are practical. On the other hand, lossy compression techniques can achieve significant reduction in data size by allowing controllable error in the reconstructed data.  Lossy  compression techniques can be further divided into methods that achieve compression by extracting and exploiting spatiotemporal correlations in the data and the methods that do not extract correlation. Examples of techniques that achieve compression by exploiting correlations in the data are  methods based on low-rank matrix and tensor decompositions, for example QR decomposition \cite{cheng2005compression}, interpolated decomposition \cite{dunton2020pass},  singular value decomposition  (SVD) \cite{therrien1992discrete,tropp2017practical} or tensor decomposition based on higher order SVD (HOSVD) \cite{KAC20,zhou2014decomposition}.  Deep learning techniques that are based on autoencoder-decoder also achieve compression by extracting nonlinear correlations in the data \cite{carlberg2019recovering,glaws2020deep,liu2019novel}. Examples of the lossy compression techniques that do not exploit correlated structure in the data are multiprecision truncation methods \cite{gong2012mloc}, mesh reduction techniques that map the data to a coarse spatiotemporal mesh. See Ref. \cite{glaws2020deep} for an overview of these techniques.  \\

The focus of this work is on correlation-exploiting compression techniques as these lossy  methods can achieve significantly more compression compared to other methods.  The performance of these techniques can be assessed by two criteria: (i) the compression ratio for a given reconstruction error, and (ii)  computational cost of performing the compression. The compression ratio is the ratio of the storage size of the full-dimensional data to that of the  compressed data. Methods that can achieve higher compression ratio for a given accuracy or achieve higher accuracy for the same compression ratio perform better on the first metric. Computational cost of performing the compression of some techniques can be exceedingly large. For example, SVD-based reductions require solving large-scale eigenvalue problems and therefore, the computational cost of SVD-based compression can be orders of magnitudes larger than advancing the simulation for one time step -- although this depends on the size of the data. This has motivated the  randomized SVD that requires a single-pass over data \cite{tropp2017practical}, thereby reducing the computational cost significantly. To reduce the computational cost, deep-learning-based compression techniques can be trained on canonical problems and then the trained network can be used in real-time for fast compression \cite{glaws2020deep}. However, it is not clear whether the pretrained model will perform well on any unseen data.  This issue of extrapolating to unseen data does not exist in SVD-based techniques, as SVD reduction does not require any training, however, SVD-based techniques often result in lower compression ratios as they can only exploit linear correlations in data as opposed to autoencoder-decoder techniques that extract nonlinear correlations.\\

%TDB part

Recently, new reduced-order modeling techniques have been introduced in which the low-rank structures are extracted directly from the model -- bypassing the need to generate data. In these techniques, a reduced-order model  (ROM) is obtained by projecting the full-order model onto a time-dependent basis (TDB).   We refer to these techniques as \emph{model-driven} ROM, as opposed to \emph{data-driven} techniques since in the model-driven formulation an evolution equation for TDB is derived. Model-driven ROMs have shown great performance in solving high-dimensional stochastic PDEs \cite{sapsis2009dynamically,cheng2013dynamically,CSK14,MN18,B19,PB20}, reduced-order modeling of time-dependent linear systems   \cite{babaee2016minimization,BMS18,donello2020computing,NBGCL21}.  TDB has also been applied  in tensor dimension reduction with applications in quantum mechanics and quantum chemistry \cite{Beck:2000aa,Bardos:2003aa,KL10,DV2020}.  \\

%   The reduced-order model (ROM) is a dimension reduction technique that can be based on the evolution of time-dependent bases (TDB). Depending on the application, the TDB evolution can be derived from the related partial differential equations (model driven reduction), or the observed data (data driven reduction). Recent ROMs, such as dynamically orthogonal (DO) \cite{sapsis2009dynamically}, bi-orthogonal (BO) \cite{cheng2013dynamically}, dynamically/bi-orthonormal (DBO) \cite{PB20}, and optimally time-dependent (OTD) \cite{babaee2016minimization} are model driven reductions with extensive applications in stochastic problems \cite{patil2021reduced,PB20} and sensitivity analysis \cite{donello2020computing}. However, in these models the derivation of TDB evolution may become hard or impossible because of the complicated governing equations, therefore, the data driven approach becomes beneficial \cite{B19} and leads to compression in our case. \\ 

The attractive feature in model-driven techniques is that the low-rank subspace is time-dependent and it adapts to changes on the full order model. In this paper, we leverage that feature and present a data-driven analog of the TDB-based compression. In particular, we present an adaptive and scalable \emph{in situ} data compression that decomposes the streaming multidimensional data using TDBs.  This technique extracts multidimensional correlations from high-dimensional streaming data. The presented method is lossy and it is adaptive, i.e., modes are added and removed to maintain the error below a prescribed threshold. Moreover, the method does not require the data history and it only utilizes the time derivative of the instantaneous data. As a result, the compression is achieved without having to solve large-scale eigenvalue  or nonconvex optimization problems. The cost of extracting the data-driven TDB grows linearly with the data size, making it suitable for large-scale simulations where only on-the-fly compression techniques are feasible. We also  present our formulation in  a versatile form, where the multidimensional data can be decomposed into arbitrarily chosen lower-dimensional bases. In this new formulation, we recover the dynamically bi-orthonormal (DBO) decomposition \cite{PB20}, and  equivalently, the dynamically orthogonal (DO) and bi-orthogonal (BO) \cite{sapsis2009dynamically,cheng2013dynamically} decompositions as special cases.    After introducing our strategy in the section \ref{sec:Method}, we test our method on four different compression problems, including the Runge function,  incompressible turbulent reactive flow, stochastic turbulent reactive flow, and three-dimensional turbulent channel flow in section \ref{sec:problem}. In section \ref{sec:Disc}, we summarize our work and suggest possible future work..\\

%\section{Methodology}\label{sec:Method}
%\input{Sections/Methods.tex}
\section{Methodology}\label{sec:Method}
 \subsection{Notations and Definitions}
We consider that streaming data are generated by a generic $d-$dimensional nonlinear time-dependent PDE expressed by:
\begin{align}
    \pfrac{v(\bm{x},t)}{t} =\mathcal{M}(v,\bm{x},t), \quad \bm{x} \in \Omega,    \quad t>0, \label{Eq:M}
\end{align}
where $\bm{x}=\{x_1,x_2, \dots, x_d\}$ are the $d$-dimensional independent variables, which include \emph{differential} dimensions  with respect to which differentiation appears in the PDE as well as \emph{parametric/random} dimensions to which the solution of the PDE has parametric dependence. The examples of the parametric space are design space or random parametric space, where different samples of parameters can be run concurrently.  In Eq. \ref{Eq:M}, $\mathcal{M}$ represents the model that in general includes linear and nonlinear differential operators augmented with appropriate boundary/initial conditions. 
% The space of independent variables is denoted by   $\Omega \subset \mathbb{R}^d$ and $v:\Omega \times \mathbb{R}^+ \rightarrow \mathbb{R}$.  
In the most generic form, we consider a disjoint decomposition of $\bm{x}$ into $p$ groups of variables: $\bm{x} = \{\bm{x}_1,\bm{x}_2, \dots, \bm{x}_p \}$, with  $2 \leq p \leq d$. The dimension of each space $\bm{x}_n$ is denoted by $d_n$. Therefore, $d=d_1+d_2+\dots+ d_p$ and the special case of  $p=d$ implies decomposing the $d$-dimensional space to $d$ one-dimensional spaces. 

% Similarly, we consider decomposing   $\Omega$ to $\Omega=\Omega^1 \times \Omega^2  \times \dots \times \Omega^p$, where $\times$ denotes  the tensorial product. \textcolor{red}{I think we should remove $\Omega$ as we discussed}% As an example, consider the   three-dimensional Navier-Stokes equations subject to 1000-dimensional uncertain boundary conditions. Here, $d=1003$ and once choice of $\bm{x}$ is $\bm{x}_1$ 
% $\tau:=\{1,2, \dots, d\}$ and $\tau^m=\{i_1,i_2, \dots, i_{d_m} \}, m=1,2, \dots, p$ where and $x^m=\{x_{i_1},x_{i_2}, \dots, x_{i_m} \}$.

In the presented methodology, we work with data, which is represented in the discrete form with vectors, matrices and tensors. However, we present our formulation using multidimensional functions, i.e., in the continuous form, since we believe this notation is easier to understand. We show how the continuous formulation can easily be turned to a discrete formulation. 

We introduce an  $L^2$ inner-product and its induced norm for the multidimensional space as in the following
% We  assume that $v$ belongs to a separable $L^2(\bm{x})$ space: $v \in L^2(\bm{x})$.  The space $L^2(\bm{x})$ is equipped with  the standard  $L^2$ inner-product and its induced norm: 
\begin{equation}\label{Eq:inner}
%  \inner{u(x_n)}{v(x_n)}_{x_n} :=  \int_{x_n}  u(x_n) v(x_n)dx_n \quad \mbox{and} \quad 
\inner{u(\bm{x})}{v(\bm{x})}_{\bm{x}} :=  \int_{\bm{x}_p}  \dots  \int_{\bm{x}_1}  u(\bm{x}) v(\bm{x})\rho(\bm{x}_1) \dots \rho(\bm{x}_p) d\bm{x}_1  \dots d\bm{x}_p  \quad \mbox{and} \quad \big \| u \big \|_{\bm{x}} := \inner{u}{u}_{\bm{x}}^\frac{1}{2},
\end{equation}
 where $\rho(\bm{x}_i)$ is the nonnegative density weight in each space.   Similarly, an inner product for $\bm{x}_n$ and its induced  $L^2$ norm can be defined as:
\begin{equation}
    \inner{u(\bm{x}_n)}{v(\bm{x}_n)}_{\bm{x}_n}= \int_{\bm{x}_n} u(\bm{x}_n)v(\bm{x}_n)\rho(\bm{x}_n) d\bm{x}_n, \quad \mbox{and} \quad \| u(\bm{x}) \|_{\bm{x}_n} = \langle u(\bm{x}),u(\bm{x}) \rangle^{\frac{1}{2}}_{\bm{x}_n}.
\end{equation} 
We also introduce the following notation for the  inner product with respect to all dimensions except $\bm{x}_j$ as follows:
 \begin{equation}\label{eq:inner_product2}
 \begin{split}
&\inner{u(\bm{x},t)}{v(\bm{x},t)}_{\bm{x} \setminus \bm{x}_j} = \\ &\int_{\bm{x}_p} \dots \int_{\bm{x}_{j+1}} \int_{\bm{x}_{j-1}}\dots  \int_{\bm{x}_1} {u(\bm{x},t)v(\bm{x},t)  \rho(\bm{x}_1) \dots \rho(\bm{x}_{j-1}) \rho(\bm{x}_{j+1}) \dots \rho(\bm{x}_p) d\bm{x}_1  \dots d\bm{x}_{j-1} d\bm{x}_{j+1}\dots d\bm{x}_p}.
\end{split}
 \end{equation}
% We also use the notation  $\inner{}{}_{\Omega \backslash \Omega^m}$ to imply the inner product in the space of $\{\bm{x}_1, \dots, \bm{x}_{m-1},\bm{x}_{m+1}, \dots, \bm{x}_p \}$, i.e., all spaces except $\Omega^m$.
% We assume that the  space of $L^2(\Omega)$ can be decomposed to:   $L^2(\Omega)= L^2(\Omega^m) \otimes L^2(\Omega^{1\dots m-1\ m+1 \dots p})$, where $L^2(\Omega^m)$ and $L^2(\Omega^{1\dots m-1\ m+1 \dots p})$ are equipped with the inner products of $\big < ~ ,  ~ \big>_{\Omega^m}$ and  $\big < ~ , ~ \big>_{\Omega \backslash \Omega^m}$, respectively. 
We also denote a set of orhtonormal time-dependent bases (TDB) in space $\bm{x}_n$ by:

\begin{equation*}
\bm{U}^{(n)}(\bm{x}_n,t)=\begin{bmatrix} u^{(n)}_1(\bm{x}_n,t) \vert u^{(n)}_2(\bm{x}_n,t) \vert \dots \vert u^{(n)}_{r_n}(\bm{x}_n,t) \end{bmatrix},
\end{equation*} 
where the superscript $(n)$ shows  the index of the dimension group and $r_n$ is the dimension of the subspace spanned by $\bm{U}^{(n)}(\bm{x}_n,t)$. We also use the tensor notation that was presented in the review article \cite{KB09}. For a $t$-order tensor $\tn{T} \in \mathbb{R}^{r_1\times r_2 \times \dots \times r_p}$,  
 the $n$th or $n$-mode unfolding of a tensor to a matrix is denoted by: $\tn{T} _{(n)}(t)  \in \mathbb{R}^{r_n \times r_{n+1} \dots r_p r_1 r_2 \dots r_{n-1}}$. The n-mode product of $\tn{T} \in \mathbb{R}^{r_1\times r_2 \times \dots r_p}$ and the matrix  $\bm{A} \in \mathbb{R}^{N_n\times r_n}$  is defined as $\tn{T} \times_n \bm{A} \in \mathbb{R}^{r_1\times r_2 \dots  \times r_{n-1}\times N_n \times r_{n+1}\times \dots  \times r_p }$. We also denote the time derivative with $\partial ( \sim )/\partial t \equiv \dot {( \sim )}$.

\subsection{Reduction via Time-dependent Bases }
We decompose the multidimensional streaming data into a time-dependent core tensor and a set of time-dependent orthonormal TDB as in the following:
  \begin{equation}\label{eq:reconstructed}
 v(\bm{x},t)= \sum_{i_{p}=1}^{r_p} \dots \sum_{i_{2}=1}^{r_2} \sum_{i_{1}=1}^{r_1}   \tn{T}_{i_1 i_2 \dots i_p}(t) u_{i_1}^{(1)} (\bm{x}_1,t) u_{i_2}^{(2)} (\bm{x}_2,t) \dots u_{i_p}^{(p)} (\bm{x}_p,t) + e (\bm{x},t),
\end{equation} 
where  $\tn{T}(t)$ is the time-dependent core tensor and $\bm{U}^{(n)}(\bm{x}_n,t)=\begin{bmatrix} u^{(n)}_1(\bm{x}_n,t) \vert u^{(n)}_2(\bm{x}_n,t) \vert \dots \vert u^{(n)}_{r_n}(\bm{x}_n,t) \end{bmatrix}$ are a set of  orthonormal modes and $e (\bm{x},t)$ is the low-rank approximation error. In the above description, $v(\bm{x},t)$ represents the streaming data, which in the discrete form  is represented by a $d$-dimensional time-dependent tensor. In the following, we derive closed-form evolution equations for the core tensor and the TDB.  For the sake of simplicity, we derive the equations for a special case with three bases,  where $p=3$. We also present the formulation for the most generic case.  The TDB is instantaneously orthonormal and therefore:
%   The inner product between two state space with respect to $\bm{x}=(x_1,x_2,\dots,x_d)$:\\
%  \begin{equation}\label{eq:inner_product1}
%  \langle u(\bm{x},t),v(\bm{x},t) \rangle_\bm{x} = \int_{x_d} \dots \int_{x_2} \int_{x_1} {u(\bm{x},t)v(\bm{x},t)  dx_1 dx_2 \dots dx_d} 
%  \end{equation}
\begin{equation}\label{eq:orhonormality}
\inner{u^{(n)}_i(\bm{x}_n,t)}{u^{(n)}_{i'}(\bm{x}_n,t)}_{\bm{x}_n}=\delta_{ii'}, \qquad i,i'=1,2,\dots,r_n,\qquad n=1,2,\dots,p .
\end{equation}
Taking time derivative of the orthonormality condition results in:
\begin{equation}\label{eq:dyna_orhonormal}
\frac{d} {d t} \inner {u^{(n)}_i(\bm{x}_n,t)}{u^{(n)}_{j} (\bm{x}_n,t)}_{\bm{x}_n}= \inner{\dot{u}^{(n)}_i(\bm{x}_n,t)}{u^{(n)}_j(\bm{x}_n,t)}_{\bm{x}_n} + \inner {u^{(n)}_i(\bm{x}_n,t)} {\dot{u}^{(n)}_j(\bm{x}_n,t) }_{\bm{x}_n} =0.
\end{equation}
Let $\bm{\phi}^{(n)}_{ij}(t) = \inner{\dot{u}^{(n)}_i(\bm{x}_n,t)}{u^{(n)}_j(\bm{x}_n,t)}_{\bm{x}_n}$, where $\bm{\phi}^{(n)}_{ij}(t) \in \mathbb{R}^{r_n\times r_n} $. From Eq.~\ref{eq:dyna_orhonormal}, it is clear that $\bm{\phi}^{(n)}_{ij}(t)$ is a skew-symmetric matrix $
\bm{\phi}^{(n)}_{ij}(t)= - {\bm{\phi}^{(n)^T}_{ij}(t)}$.
Based on the above definitions, we derive the evolution equations for the bases and the core tensor. To this end, we take a time derivative of the TDB decomposition given by Eq. \ref{eq:reconstructed}. This follows:
\begin{equation} \label{eq:ex1Ddot}
\dot{v}  = \tn{\dot{T}} _{i_1 i_2 i_3}  u_{i_1}^{(1)} u_{i_2}^{(2)} u_{i_3}^{(3)} + \tn{T} _{i_1 i_2 i_3} \dot{u}_{i_1}^{(1)} u_{i_2}^{(2)} u_{i_3}^{(3)}  + \tn{T} _{i_1 i_2 i_3} u_{i_1}^{(1)} \dot{u}_{i_2}^{(2)} u_{i_3}^{(3)} + \tn{T} _{i_1 i_2 i_3} u_{i_1}^{(1)} u_{i_2}^{(2)} \dot{u}_{i_3}^{(3)}.
\end{equation}
By taking the inner product $\inner {\bullet} {u_{i'_1}^{(1)} u_{i'_2}^{(2)} u_{i'_3}^{(3)}}_{\bm{x}}$ of  Eq. \ref{eq:ex1Ddot} and using orthonormality conditions, the evolution of the core tensor is obtained as follows:
\begin{equation} \label{eq:ex2Core}
\tn{\dot{T}}_{i'_1 i'_2 i'_3}  = \inner{ \dot{v}} {u_{i'_1}^{(1)} u_{i'_2}^{(2)} u_{i'_3}^{(3)}}_{\bm{x}} - \tn{T} _{i_1 i'_2 i'_3} \phi_{i'_1 i_1}^{(1)} - \tn{T} _{i'_1 i_2 i'_3} \phi_{i'_2 i_2}^{(2)} - \tn{T} _{i'_1 i'_2 i_3} \phi_{i'_3 i_3}^{(3)}.
\end{equation}
To derive the evolution equation for the bases, we start with deriving an expression for $\dot{u}^{(1)}_{i_1}$ by taking the inner product $\inner {\bullet} {u_{i'_2}^{(2)} u_{i'_3}^{(3)}}_{\bm{x} \setminus \bm{x}_1}$  of Eq. \ref{eq:ex1Ddot} as follows:
\begin{equation}\label{eq:ex3}
\inner{\dot{v}}{ u_{i'_2}^{(2)} u_{i'_3}^{(3)}}_{\bm{x} \setminus \bm{x}_1}  = \tn{\dot{T}} _{i_1 i'_2 i'_3}  u_{i_1}^{(1)} + \tn{T} _{i_1 i'_2 i'_3} \dot{u}_{i_{1}}^{(1)} + 
\tn{T} _{i_1 i_2 i'_3} u_{i_{1}}^{(1)} \phi_{i'_2 i_2}^{(2)} + \tn{T} _{i_1 i'_2 i_3} u_{i_{1}}^{(1)} \phi_{i'_3 i_3}^{(3)}.
\end{equation}
By rearranging the indexes for $\tn{\dot{T}}_{i'_1 i'_2 i'_3}$ from Eq. \ref{eq:ex2Core}:
\begin{equation*}
\tn{\dot{T}}_{i_1 i'_2 i'_3}  = \inner{ \dot{v}} {u_{i_1}^{(1)} u_{i'_2}^{(2)} u_{i'_3}^{(3)}}_{\bm{x}} - \tn{T} _{j_1 i'_2 i'_3} \phi_{i'_1 j_1}^{(1)} - \tn{T} _{i'_1 i_2 i'_3} \phi_{i'_2 i_2}^{(2)} - \tn{T} _{i'_1 i'_2 i_3} \phi_{i'_3 i_3}^{(3)}.
\end{equation*}
We can substitute it into Eq.~\ref{eq:ex3} and derive the evolution equation for the first bases:
\begin{equation}\label{eq:Udot1}
\tn{T}_{i_1 i'_2 i'_3} \dot{u}_{i_{1}}^{(1)} = \inner{ \dot{v}}{u_{i'_2}^{(2)} u_{i'_3}^{(3)} }_{\bm{x} \setminus \bm{x}_1} - u_{i_{1}}^{(1)}[\inner{ \dot{v}} {u_{i_1}^{(1)} u_{i'_2}^{(2)} u_{i'_3}^{(3)}}_{\bm{x}}  -\tn{T}_{j_1 i'_2 i'_3} \phi_{i_1 j_1}^{(1)}].
\end{equation}
We denote the orthogonal projection onto the complement space spanned by $\bm{U}^{(1)}$ as in the following:
\begin{equation*}
    \underset{ {\bm{U}}^{(1)}}{\prod} \inner{ \dot{v}}{u_{i'_2}^{(2)} u_{i'_3}^{(3)} }_{\bm{x} \setminus \bm{x}_1}= \inner{ \dot{v}}{u_{i'_2}^{(2)} u_{i'_3}^{(3)} }_{\bm{x} \setminus \bm{x}_1} - \inner {\inner{ \dot{v}}{u_{i'_2}^{(2)} u_{i'_3}^{(3)} }_{\bm{x} \setminus \bm{x}_1}}{u_{i'_1}^{(1)}} u_{i_1}^{(1)}.
\end{equation*}
Incorporating the above definition into  Eq. \ref{eq:Udot1} and rearranging the equation results in:
\begin{equation*}
\tn{T} _{i_1 i'_2 i'_3} \dot{u}_{i_{1}}^{(1)} =  \underset{ \bm{U}^{(1)}}{\prod} \inner{ \dot{v}}{u_{i'_2}^{(2)} u_{i'_3}^{(3)} }_{\bm{x} \setminus \bm{x}_1} + \tn{T} _{i_1 i'_2 i'_3} \phi_{i'_1 i_1}^{(1)} u_{i_1}^{(1)}.
\end{equation*}
The above equation can be an underdetermined system with respect to unknowns $\dot{u}_{i_{1}}^{(1)}$ if $r_1>r_2 r_3$ and it could be overdetermined otherwise. In order to address this issue, we find the least-square solution for $\dot{u}_{i_{1}}^{(1)}$. This can be accomplished by first multiplying both sides of the above equation by $\tn{T} _{(1)}^T$ and then computing the inverse of the resulting matrix. This amounts to computing the  pseudoinverse of the unfolded core tensor, which is denoted  by $\bm{T} ^{(1) ^ \dagger}$. The resulting equation is: %\hb{all tensor unfolding indices must be put in parentheses }
\begin{equation*}
 \dot{u}_{i_{1}}^{(1)}  =   \underset{ \bm{U}^{(1)}}{\prod} \inner{ \dot{v}}{u_{i'_2}^{(2)} u_{i'_3}^{(3)} }_{\bm{x} \setminus \bm{x}_1} \bm{T}^{(1) ^ \dagger}_{i_1,i'_2 i'_3}  + u_{i'_{1}}^{(1)} \phi_{i'_1 i_1}^{(1)}, \quad \bm{T} ^{(1) ^ \dagger} = \tn{T} _{(1)}^T (\tn{T} _{(1)}\tn{T} _{(1)}^T)^{-1}, \quad \bm{T} ^{(1) ^ \dagger}_{i_1,i'_2 i'_3} = \bm{T}^{(1)^ \dagger} _{i_1, i'_2+(i'_3-1)r_2}.
\end{equation*}
In a similar manner, we can derive $\dot{u}_{i_{2}}^{(2)}$ and $\dot{u}_{i_{3}}^{(3)}$:
\begin{equation*}
 \dot{u}_{i_{2}}^{(2)}  =   \underset{ \bm{U}^{(2)}}{\prod} \inner{ \dot{v}}{u_{i'_1}^{(1)} u_{i'_3}^{(3)} }_{\bm{x} \setminus \bm{x}_2} \bm{T} ^{(2) ^ \dagger} _{i_2,i'_1 i'_3}  + u_{i'_{2}}^{(2)} \phi_{i'_2 i_2}^{(2)}, \quad \bm{T} ^{(2) ^ \dagger}= \tn{T} _{(2)}^T (\tn{T} _{(2)}\tn{T} _{(2)}^T)^{-1}, \quad \bm{T}^{(2) ^ \dagger}_{i_2,i'_1 i'_3} = \bm{T} ^{(2) ^ \dagger} _{i_2, i'_1+(i'_3-1)r_1},
\end{equation*}
\begin{equation*}
 \dot{u}_{i_{3}}^{(3)}  = \underset{ \bm{U}^{(3)}}{\prod} \inner{ \dot{v}}{u_{i'_1}^{(1)} u_{i'_2}^{(2)} }_{\bm{x} \setminus \bm{x}_3} \bm{T} ^{(3) ^ \dagger}_{i_3,i'_1 i'_2} + u_{i'_{3}}^{(3)} \phi_{i'_3 i_3}^{(3)}, \quad \bm{T} ^{(3) ^ \dagger} = \tn{T} _{(3)}^T (\tn{T} _{(3)}\tn{T} _{(3)}^T)^{-1}, \quad \bm{T} ^{(3) ^ \dagger}_{i_3,i'_1 i'_2} = \bm{T} ^{(3) ^ \dagger} _{i_3, i'_1+(i'_2-1)r_1}.
\end{equation*}
Similarly, we can derive the evolution equations for the general case where $2 \leq p \leq d$. To this end, we take the time derivative of Eq. \ref{eq:reconstructed} as follow:  
\begin{equation}\label{eq:Ddot}
\begin{split}
& \dot{v}  = \tn{\dot{T}} _{i_1 i_2 \dots i_p}  u_{i_1}^{(1)} u_{i_2}^{(2)} \dots u_{i_p}^{(p)} + \tn{T} _{i_1 i_2 \dots i_p} \dot{u}_{i_1}^{(1)} u_{i_2}^{(2)} \dots u_{i_p}^{(p)}  + \tn{T} _{i_1 i_2 \dots i_p} u_{i_1}^{(1)} \dot{u}_{i_2}^{(2)} \dots u_{i_p}^{(p)} + \dots  \\ & +  \tn{T} _{i_1 i_2 \dots i_p} u_{i_1}^{(1)} u_{i_2}^{(2)} \dots \dot{u}_{i_p}^{(p)}.
\end{split}
\end{equation}
By taking the inner product $\inner {\bullet} {u_{i'_1}^{(1)} u_{i'_2}^{(2)} \dots u_{i'_p}^{(p)}}_{\bm{x}}$ of both sides of the Eq. \ref{eq:Ddot} and using orthonormality conditions, the evolution of core tensor is obtained   as follows:
\begin{equation}\label{eq:Core1}
 \tn{\dot{T}}_{i'_1 i'_2  \dots i'_p}  = \inner{ \dot{v}} {u_{i'_1}^{(1)} u_{i'_2}^{(2)} \dots u_{i'_p}^{(p)}}_{\bm{x}} - \tn{T} _{i_1 i'_2 \dots i'_p} \phi_{i'_1 i_1}^{(1)} - \tn{T} _{i'_1 i_2 \dots i'_p} \phi_{i'_2 i_2}^{(2)} - \dots  - \tn{T} _{i'_1 i'_2 \dots i_p} \phi_{i'_p i_p}^{(p)}.
\end{equation}
To derive the evolution equation $\dot{u}_{i_j}(\bm{x}_j,t)$, we project both sides of Eq. \ref{eq:Ddot} onto all bases except $u_{i_j}(\bm{x}_j,t)$. This can be accomplished by taking the inner product  $ \inner{\bullet}{ u_{i'_1}^{(1)} u_{i'_2}^{(2)} \dots u_{i'_{j-1}}^{(j-1)}  u_{i'_{j+1}}^{(j+1)}  \dots  u_{i'_p}^{(p)}  }_{\bm{x} \setminus \bm{x}_j}$ of  Eq. \ref{eq:Ddot}. Similar to the simplified derivation for $p=3$, we substitute the evolution of the core tensor from Eq. \ref{eq:Core1} into our operations. This yields to:\\
\begin{equation}\label{eq:basis1}
\dot{u}_{i_{j}}^{(j)}  = \underset{ \bm{U}^{(j)}}{\prod} \inner{ \dot{v}}{u_{i'_1}^{(1)} u_{i'_2}^{(2)} \dots u_{i'_{j-1}}^{(j-1)} u_{i'_{j+1}}^{(j+1)} \dots u_{i'_p}^{(p)}   }_{\bm{x} \setminus \bm{x}_j} \bm{T}^{(j)^ \dagger}_{i_j,i'_1 i'_2 \dots i'_{j-1} i'_{j+1} \dots i'_{p}} + u_{i_{j}}^{(j)} \phi_{i'_j i_j}^{(j)}.%, \quad \bm{T} _{j} ^ \dagger = \tn{T} _{j}^T (\tn{T} _{j}\tn{T} _{j}^T)^{-1}
\end{equation}
%In the appendix we  show
Here, any skew symmetric choice for $\phi$ results in an equivalent TDB approximation.  For simplicity, we can choose it to be zero. Therefore, Eq. \ref{eq:Core1} and \ref{eq:basis1} would become as follow:
\begin{equation}
 \tn{\dot{T}}_{i'_1 i'_2  \dots i'_p}  = \inner{ \dot{v}} {u_{i'_1}^{(1)} u_{i'_2}^{(2)} \dots u_{i'_p}^{(p)}}_{\bm{x}}, \label{eq:Corecont}
\end{equation}
\begin{equation}
 \dot{u}_{i_{j}}^{(j)}  = \underset{ \bm{U}^{(j)}}{\prod} \inner{ \dot{v}}{u_{i'_1}^{(1)} u_{i'_2}^{(2)} \dots u_{i'_{j-1}}^{(j-1)} u_{i'_{j+1}}^{(j+1)} \dots u_{i'_p}^{(p)}   }_{\bm{x} \setminus \bm{x}_j} \bm{T}^{(j)^ \dagger}_{i_j,i'_1 i'_2 \dots i'_{j-1} i'_{j+1} \dots i'_{p}}. \label{eq:basecont}
\end{equation}
Not only does the choice of $\phi=0$ simplify the above evolution equation but also implies a simple interpretation where $\dot{u}^{(j)}_{i_j}$ is always orthogonal to the space spanned by $u^{(j)}_{i_j}$ (dynamically orthogonal condition). This choice is also used for reduced-order modeling in Ref. \cite{B19,PB20,donello2020computing,RNB21}.
%\hb{See my observation-driven paper and highlight key points about the TDB evolution equations.  In particular, look at what I wrote after eqn 2.12 in that paper.}

Equations ~\ref{eq:Corecont} and \ref{eq:basecont} constitute the evolution equations for the TDB compression scheme. We note the DBO decomposition \cite{PB20,RNB21} is a special case of the above scheme where $p=2$. It has also been shown in Ref. \cite{PB20} that DO and BO decompositions are equivalent to DBO, i.e., DO, BO and DBO extract the same subspace from the full-dimensional problem. In that sense, DO and BO are closely related to Eqs.~\ref{eq:Corecont} and \ref{eq:basecont} for the special case of $p=2$. It is possible to derive data-driven evolution in the DO and BO forms.   The data-driven formulation of the DO decomposition is presented in Ref. \cite{B19}.

Equations ~\ref{eq:Corecont} and \ref{eq:basecont} must be solved in the space-time discrete form and   they must be augmented with appropriate initial conditions.  

\subsection{Time-Dependent Bases in Discrete Space-Time Form}
\begin{figure}[t!]
\centering
\includegraphics[width=0.8\textwidth]{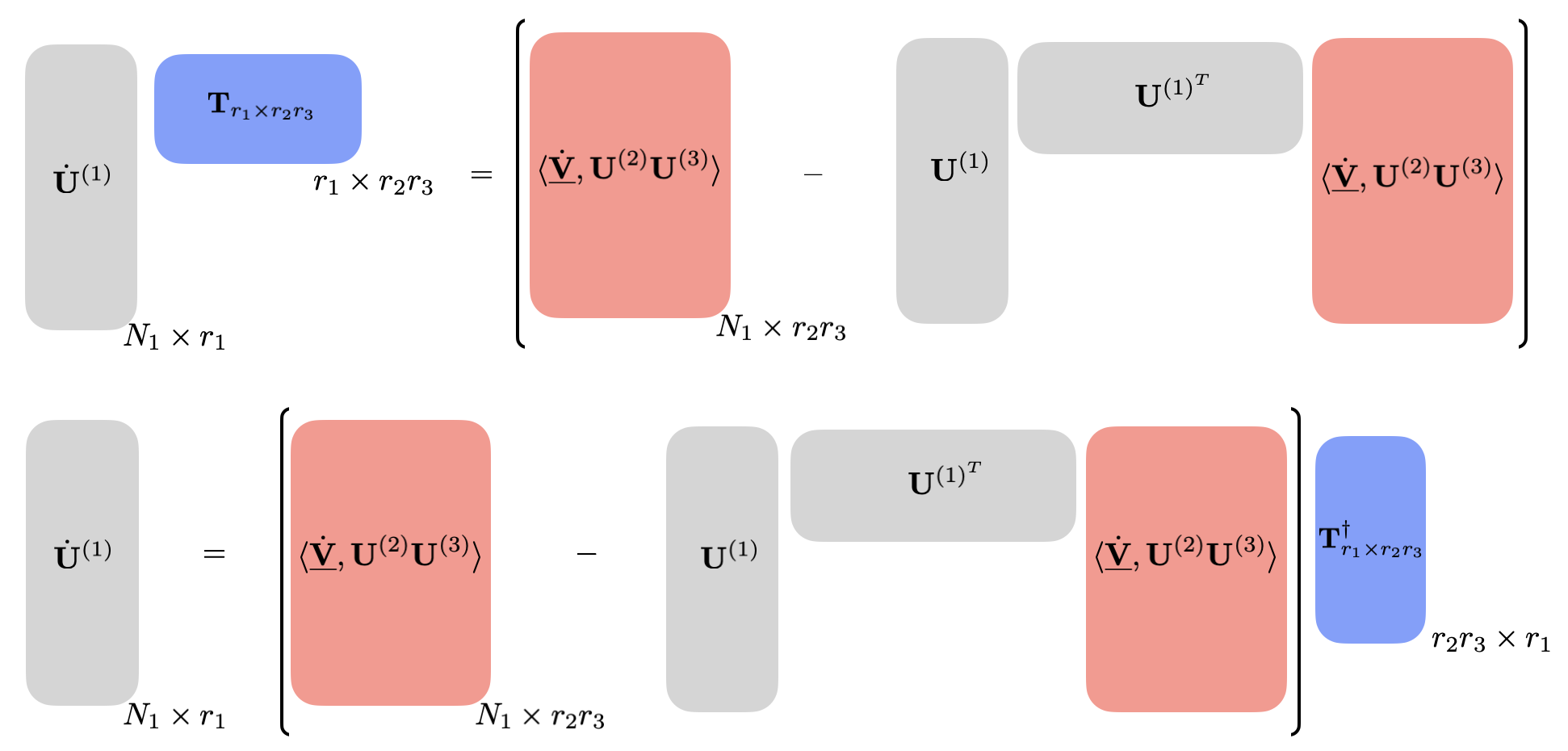}
\caption{Time-dependent bases in discrete form. For simplicity weights are considered to be identity $(\bm{W}_{N_1 \times N_1}^{(1)} = \bm{I})$.}
\label{Fig:discrete}
\end{figure}

In the previous section, we presented the TDB decomposition in the continuous space-time form. In this section, we show the details of how the above scheme can be applied to streaming data that are discrete in space and time.  In the space-discrete form, the multidimensional data ($v(\bm{x},t)$) are represented by a $p$-order time-dependent tensor denoted by: $\tn{V}(t) \in \mathbb{R}^{N_1 \times N_2 \times \dots \times N_p}$. If data are generated by a simulation,  $N_1,N_2,\dots, N_p$ are the number of grid points. Note that the data may be  generated by higher-dimensional independent  variables, i.e., $d\geq p$.   The orthonormal bases in the $\bm{x}_n$ dimension can be represented as a time-dependent matrix: $\bm{U}^{(n)}(t) =[\bm{u}^{(n)}_1(t), \bm{u}^{(n)}_2(t), \dots, \bm{u}^{(n)}_{r_n}(t) ] \in \mathbb{R}^{N_n \times r_n}$, where $\bm{u}^{(n)}_i$ is a discrete (vector) representation of $u_i$. We show the discrete form of the TDB formulation in figure \ref{Fig:discrete} where the inner product in the space is approximated with a quadrature rule: 
\begin{equation}
    \inner{u_i^{(n)}}{u_j^{(n)}}_{\bm{x}_n} \simeq {\bm{u}_i^{(n)}}^T \bm{W}^{(n)} \bm{u}_j^{(n)}, 
\end{equation}
where $\bm{W}^{(n)} \in \mathbb{R}^{N_n \times N_n}$
 is the diagonal matrix, whose diagonal elements are the quadrature weights.  The time derivative of the streaming data  ($\tn{\dot{V}}(t)$) is computed with finite difference  and the evolution equation for TDB and the core tensor can be  advanced with a standard time-integration scheme.    Various time discretization schemes can be used. High-order finite-difference discretizations  require keeping the solution from  multiple time steps in the memory.   In the first demonstration example, we investigate the error introduced by using different temporal schemes for both $\tn{\dot{V}}(t)$ and the evolution of TDB. 
 
 As we discuss in the next section, the TDB decomposition and the instantaneous HOSVD are closely connected.  To this end, HOSVD of the full-dimensional data at $t=0$ is used to initialize the core tensor as well as the TDB.

\subsection{Extraction of Coherent Structure from Streaming Data}\label{sec:CS}
%One of the key advantages of the presented methodology is the interpretability of the \\
The TDB decomposition is closely related to the instantaneous HOSVD of the multidimensional data. The HOSVD extracts correlated structures for all $p$ unfoldings of tensor $\tn{V}(t)$. As we demonstrate in our examples, the  HOSVD of the TDB core tensor $\tn{T}(t)$ follows the  leading singular values obtained by HOSVD of the full tensor $\tn{V}(t)$. To establish the connection between HOSVD and TDB,    let 
\begin{equation}
  \tn{T}_{(n)}(t)= {\bs{\Psi}^{(n)}}(t) \bm{\Sigma}^{(n)}(t) {\bs{\Theta}^{(n)^T}}(t),   
\end{equation}
 be the SVD of the unfolded core tensor, where  $\bm{\Sigma}^{(n)} =\mbox{diag}(\sigma^{(n)}_1, \dots, \sigma^{(n)}_{r_n})$ are the singular values of the $n$-mode unfolding of the core tensor, $\bs{\Psi}^{(n)}(t)$ and  $\bs{\Theta}^{(n)}(t)$ are the left and right singular vectors of the unfolded core tensor.  As we show in our demonstrations, $\bm{\Sigma}^{(n)}$ closely follows the $r_n$  leading singular values of the  $n$-mode unfolding of the full-dimensional streaming  data  ($\tn{V}_{(n)}(t)$). Moreover, the TDB closely follows the leading left singular vectors of $\tn{V}_{(n)}(t)$  after rotating TDB along the energetically ranked direction by using $\bs{\Psi}^{(n)}(t)$ as in the following:
 \begin{equation}\label{eq:rot}
     \tilde{\bm{U}}^{(n)}(t)=\bm{U}^{(n)}(t) \bs{\Psi}^{(n)}(t).
 \end{equation}
 In other words, the TDB closely approximate the same subspace  spanned by the leading left singular vectors of the the unfolded data. With the rotation given by Eq. \ref{eq:rot}, the TDB modes are ranked energetically and they can be compared against the left singular vectors of the the unfolded data. In the case of $p=2$, where the high-dimensional data are matricized  and TDB  reduces to DBO,   agreements  between the TDB subspace and the instantaneously optimal subspace obtained from SVD of the full-dimensional data have already been established. See Ref. \cite{PB20} for the case of stochastic reduced-order modeling, Ref. \cite{RNB21} for the case of reduced-order modeling of reactive species transport equation. This is also true for DO \cite{SL09}, OTD \cite{Babaee_PRSA} and BO \cite{CHZI13} formulations, which are all equivalent to the DBO formulation. In the case of linear dynamics, the convergence of TDB modes to the dominant subspace obtained by the SVD of the full-dimensional data is theoretically shown \cite{Babaee_PRSA,BFHS17}.

 The $\tilde{\bm{U}}^{(n)}(t)$ modes obtained from Eq. \ref{eq:rot} captures the dominant structures among the columns of the $n$th unfolding of the full-dimensional   data, and therefore,   $\tilde{\bm{U}}^{(n)}(t)$ can be interpreted as instantaneous coherent structures in the streaming data. 
%  In this method, we can calculate singular values in a direction by taking singular value decomposition (SVD) from the unfolded core tensor in the same direction $\tn{T}_n= {\bm{R}} \bm{\Sigma}^{(n)} {\bm{Q}} $, where $\bm{\Sigma}^{(n)}$ is a diagonal matrix with ranked singular values (i.e $\bm{\sigma}_{x_n}=diag(\bm{\Sigma}^n)$). Here, $\bm{R}$ and $\bm{Q}$ are orthonormal matrices and $\bm{R}$ can be used to rotate TDB modes $\tilde{\bm{U}}^{(n)}=\bm{U}^{(n)} \times \bm{R}$ and make them equivalent to factor matrices from HOSVD $\tilde{\bm{U}}^{(n)}$. HOSVD singular values calculation is the same as TDB. 
 %HOSVD process  \\
%present them in canonical format. 
%\hb{Give more details on HOSVD and its connection to TDB}XXX

\subsection{Error Control and Adaptivity}
For highly transient systems, the rank of the systems may change as the system evolves. Therefore, to maintain the error at a desirable level, the multirank $(r_1,r_2 ,\dots, r_p)$ must change in time accordingly. There are two types of error in TDB decomposition: (i) temporal discretization error, and (ii) the error of the unresolved subspace, which are a result of neglecting the interactions of the resolved TDB subspace with the unresolved subspace. The lost interactions induce  a \emph{memory error} on TDB components that can be properly analyzed in the Mori-Zwanzig framework \cite{Chorin28032000}. Unlike the model-driven TDB,  in the data-driven mode, the error can be monitored by computing the Frobenius norm of the difference between the full data and TDB reconstruction, i.e.,
 \begin{equation}\label{eq:error}
 \varepsilon(t)=\big\|\tn{V}(t)-\tn{V}^{TDB}(t) \big\|_F.
 \end{equation}
 The above Frobenius norm is defined based on the weighted inner product. 
 To develop an adaptive strategy for TDB, we need to define an error criterion for mode addition/removal. 
 Using singular values, we can calculate the percentage of the captured detail as follow:
 \begin{equation}
 \gamma^{(n)}(r)= \frac{\Sigma_{i=1}^{r_n} \sigma_{i}^{(n)^2}}{ \Sigma_{i=1}^{N_n} \sigma_{i}^{(n)^2}} \times 100 = \frac{\Sigma_{i=1}^{r_n} \sigma_{i}^{(n)^2}}{\Sigma_{i=1}^{r_n} \sigma_{i}^{(n)^2} + \varepsilon^2 (t)} \times 100.
 \label{eq:percentage}
\end{equation}
% where error $\varepsilon (t)$ is based on the $L^2$ norm of the difference between reconstructed TDB and the streaming data (i.e. $\varepsilon(t)=\|\tn{V}(t)-\tn{V}^{TDB}(t) \|_2$) and is equal to $\sqrt{\Sigma_{i=r+1}^{N} \sigma_{i_{x_n}}^2}$.
 
\begin{figure}[t!]
\centering
\includegraphics[width=0.9\textwidth]{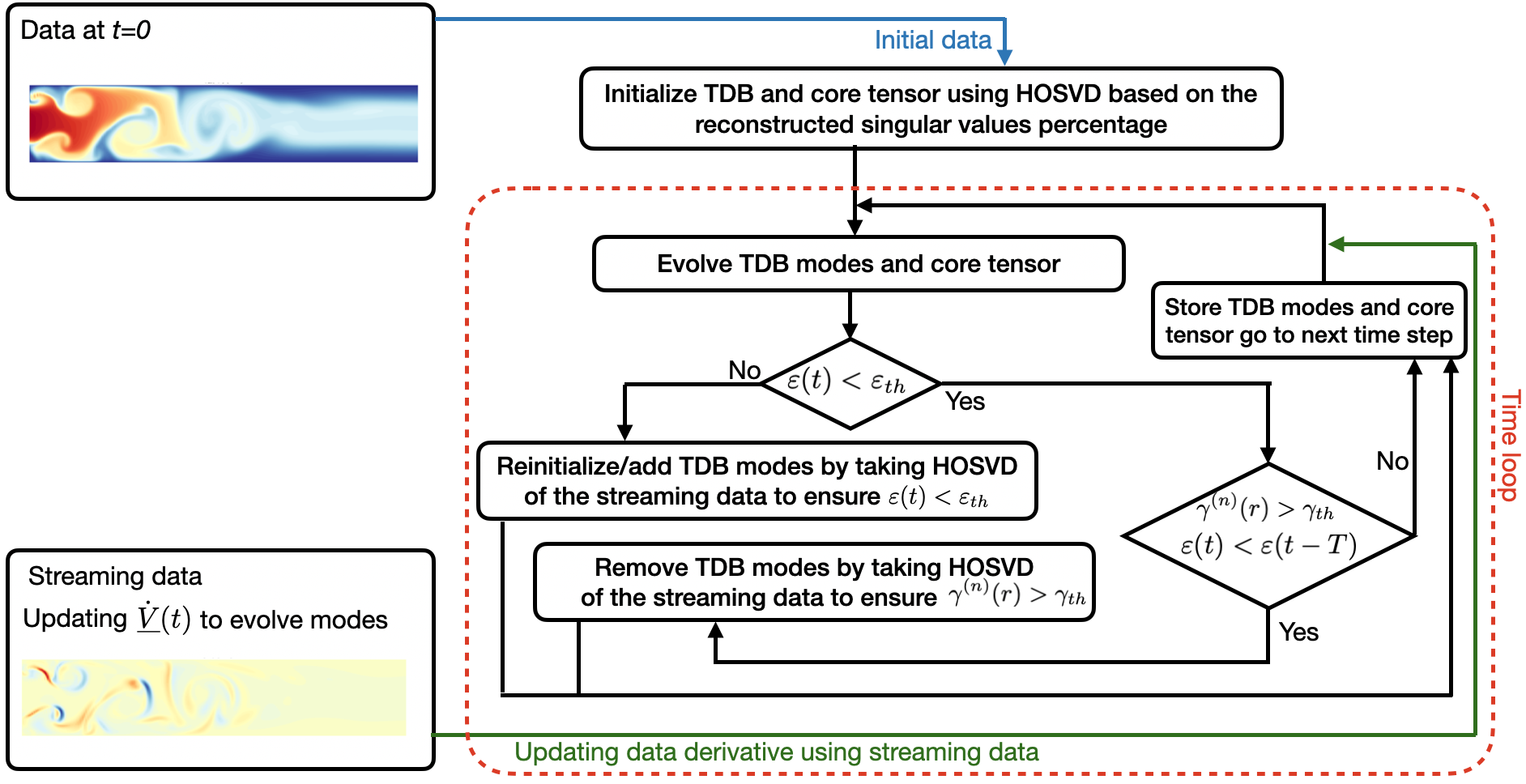}
\caption{Algorithm flowchart for the compression and error adaptivity.}
\label{Fig:Algorithm}
\end{figure}
In order to control the error, we define a maximum allowable error for $ \varepsilon(t)$, which is chosen by the practitioner. The adaptive algorithm  adds  modes if the error exceeds the threshold value and removes modes if the error can be maintained below the threshold with lower ranks based on the defined $\gamma^{(n)} (r)$. In particular, if the error exceeds the maximum error ($\varepsilon_{th}$), the algorithm: (i) reinitializes the core tensor and TDB; (ii) increases the number of modes if the resulting error from reinitialization stays above the limit; (iii) the algorithm reinitializes and reduces the number of modes if the solution starts to capture unnecessary energy and has a negative average slope for a defined number of iterations. In this process, we can increase/decrease the number of modes with respect to the relation between the defined error and singular values Eq. (\ref{eq:percentage}). For example, assume the maximum error limit is set to be $\varepsilon(t)= \varepsilon_{th}$. First, we initialize the TDB and the core tensor using HOSVD and calculate the required number of modes in each direction based on $\gamma^{(n)} (r)= \gamma_{th}$. During the computation, if $\varepsilon(t) > \varepsilon_{th}$, the algorithm reinitializes the modes. If the resulting error is not less than the prescribed limit, the algorithm increases the number of modes based on $\gamma_{th}$. In the subsequent  time steps if $\gamma(t) > \gamma_{th}$, the algorithm decreases the number of modes. Note that, rank adjustment reinitializes the modes and the core tensor, which may reduce the error for only a few iterations. This causes frequent rank adjustment with high computational costs. Therefore, in addition to the defined error, the slope of the error for a defined number of iterations (excluding the time steps with HOSVD reinitialization) must be negative to trigger the rank reduction process. We show the algorithm of this example in figure \ref{Fig:Algorithm}.
 % calculate the number of modes based on If $\sigma_{x_n}$ are resulting singular value from the unfolded core tensor $\tn{T}_{n}$,
%$e(r)= 1 - \frac{\Sigma_{i=1}^{r_{n}} \sigma_{\bm{x}_n}^2}{\Sigma_{i=1}^p \sigma_{\bm{x}_n}^2}$ describes unresolved resolution. The

%This strategy is shown in figure \ref{Fig:SVdemo1} for three-dimensional Runge function problem in section \ref{sec:problem}. In this problem an error buffer region is set to be in the interval of $\epsilon \in [5\times10^{-7},  10^{-6}]$. \hb{You should describe the method generically  and not for a specific threshold value.} This error tolerance is set by the practitioner.  It is evident that errors of the  fixed-rank approximations $r=2$ and 3 exceed the upper limit of the $10^{-6}$, while the adaptive TDB maintain the error below the upper limit by increasing the rank to $r=4$ and $r=5$. The number of modes is later reduced to $r=4$ as the dimensionality of the problem decreases for the given error.

 %For example, figure \ref{Fig:Errororder} shows the effect of both unresolved and numerical errors for a known function that changes in time. In figure \ref{Fig:errorO} high order singular value decomposition (HOSVD) error shows the unresolved error. Clearly, the first and second order Euler (EE) and RK scheme add numerical error. Here, the second order RK scheme has less error. This figure also shows the error difference between the exact data derivative, first order EE, and second order RK discretization.\\
 
\subsection{Scalability and Compression Ratio}
\begin{figure}[t!]
\centering
\subfigure[]{
\includegraphics[width=0.4\textwidth]{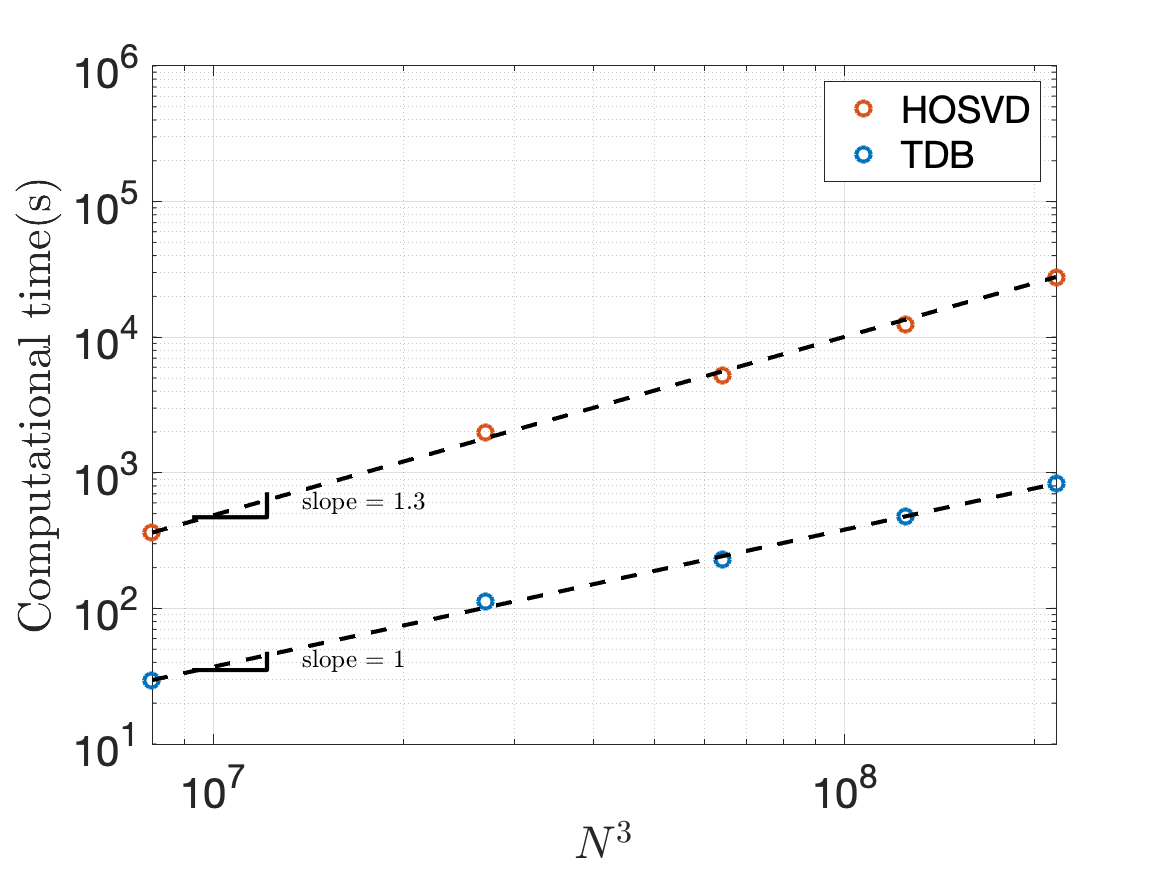}
\label{Fig:compt}
}
\subfigure[]{
\includegraphics[width=0.4\textwidth]{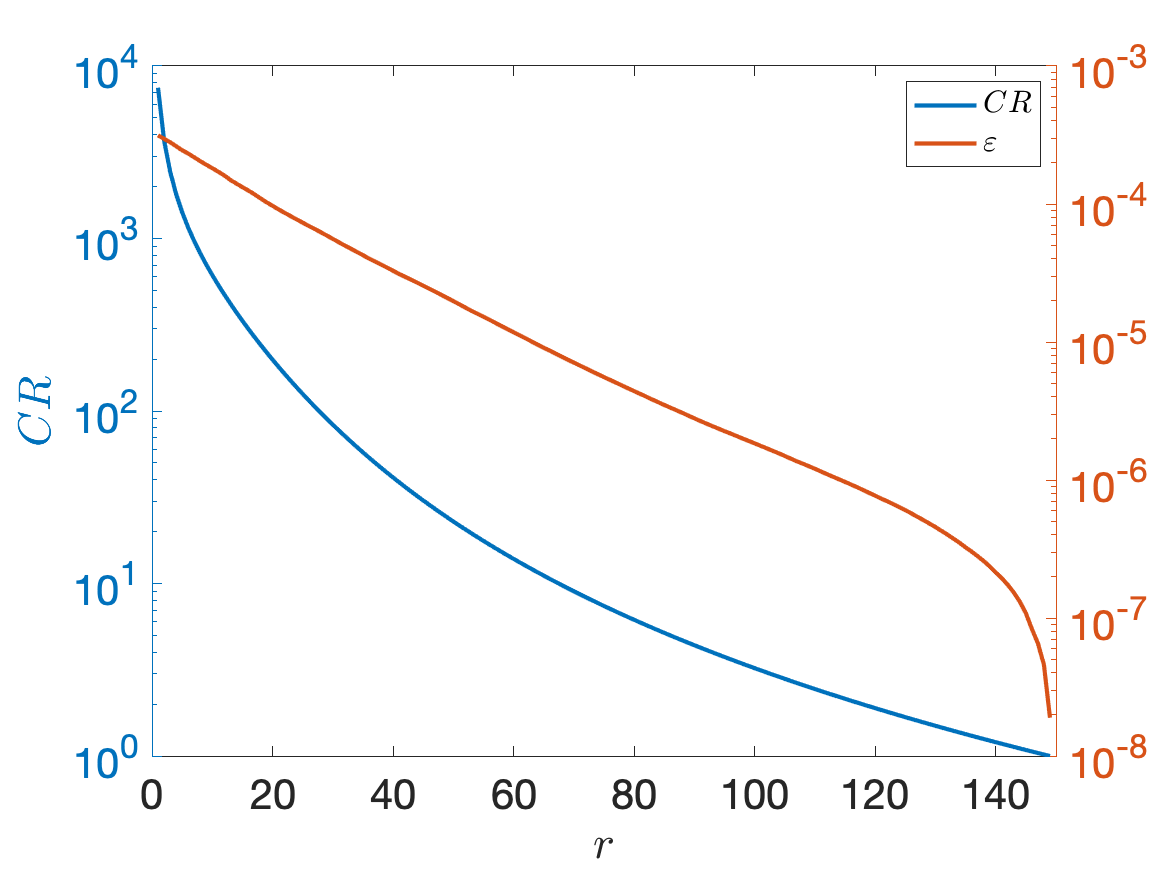}
\label{Fig:compt2}
}
\caption{Computational performance for turbulent channel flow: (a) Scalability, (b) Compression ratio and resulting error.}
\label{Fig:Computation}
\end{figure}

We show that the computational complexity of solving the resulting equations  linearly scales with the size of the data. Here, for simplicity, we consider ($p=d$), i.e., one-dimensional TDB and $N=N_1=N_2=\dots=N_d$. In this case, the total size of the data is $S=N^d$. We also consider the case where the core tensor has  equal $n$-ranks, i.e., $r=r_1=r_2=\dots=r_d$.  The leading costs of evolving TDB equations are: 
\begin{enumerate}
    \item Projection of the time-derivative of the data onto TDB ($\bm{U}^{(n)^T}  \bm{\dot{\tn{V}}}_{(n)}$), which is $\mathcal{O}(r d S)$.
    \item Computing the pseudoinverse of the unfolded core tensor ($\bm{T}^{(n) ^ \dagger}$ ). This requires the computation of  $\tn{T} _{(n)}\tn{T} _{(n)}^T$, which scales with  $\mathcal{O}(r^d)$ and the computation of the inverse of this matrix  $(\tn{T} _{1}\tn{T} _{1}^T)^{-1})$, which scales with $\mathcal{O}(r^3)$.
    \end{enumerate}  
    We make the following observations about the cost of evolving TDB: \begin{enumerate}
    \item When $r << N$, the computational cost of computing the pseudoinverse of the unfolded core tensor is negligible to the projection of the time-derivative of the data onto TDB, which scales linearly with the total size of the data. 
    \item The evolution of TDB components does not require computation of any eigenvalue problem as required in SVD-based reductions or solving a nonlinear optimization problem as is required in autoencoder-decoder reductions \cite{liu2021high,glaws2020deep,mishra2020wavelet}.
    \item Many entries of $\dot{\tn{V}}$ have small values. See figure \ref{Fig:Algorithm} for an example of $\dot{\tn{V}}$.  Therefore, the sparse approximation of $\dot{\tn{V}}$ can significantly reduce overall cost, although this advantage of TDB has not been explored in this work.
    \end{enumerate}
    The computational cost of solving the TDB equations can be compared against that of computing  HOSVD. HOSVD requires computing the SVDs of the unfolded tensor $\tn{V}_{(n)}$ for $n=1,2, \dots, d$. This requires computation and storage of the correlation matrix $\bm{C}^{(n)} = \tn{V}_{(n)}^T \tn{V}_{(n)}$, which scales with $\mathcal{O}(dN^{d+1})$ and the eigenvalue computation of this matrix, which scales with $N^3$.  The computational cost of TDB and HOSVD for a block of data generated by the turbulent channel flow  simulation (last demonstration) for the case of $N=N_1=N_2=N_3$  are shown in figure \ref{Fig:compt}. This confirms that the computational cost of TDB scales linearly with the total data size $S=N^3$, while the computational cost of  HOSVD scales with $N^3 = S^{4/3}$. This shows that TDB computations are significantly faster than HOSVD  and the disparity between the computational cost of TDB and HOSVD only grows as the dimension or number of grid points (or samples) increases.
    
  The compression ratio of the TDB decomposition is computed as the ratio of the storage cost of storing the full-dimensional data to that of solving the TDB components. The TDB  requires storing the core tensor and the orthonormal lower-dimensional bases. Therefore, the compression ratio for TDB is:
\begin{equation*}
    CR= \frac{N_1 N_2 \dots N_p}{r_1 N_1 + r_2 N_2 + \dots + r_p N_p+ r_1 r_2 \dots r_p}.
\end{equation*}
Figure \ref{Fig:compt2} shows the compression ratio ($CR$) for the same problem with the corresponding error resulting from dimension reduction. Since our algorithm is adaptive, it can change the compression ratio by changing the number of modes; therefore, we introduce the weighted compression ratio $\overline{CR}$ as follow: 
\begin{equation}
    \overline{CR}= \frac{t_m-t_0}{ \frac{(t_1 - t_0)}{CR_1} + \frac{(t_2 - t_1)}{CR_2} \dots + \frac{(t_m - t_{m-1})}{CR_m}}.
\end{equation}
Where, $CR_1, CR_2, \dots CR_m$ are compression ratios for $(t_1 - t_0),(t_2 - t_1),\dots,(t_m - t_{m-1})$ time intervals, respectively. This equation allows us to measure the effective compression ratio for all time intervals when the number of modes changes in adaptivity process.  

%\subsubsection*{\textbf{Generalized format for multi-dimensional TDB}}
%\textcolor{red}{XXXX}

\section{Demonstration Cases}\label{sec:problem}

%  bullet point:\\
%  1) Provide the Rung function for error Analysis\\
%  2) Provide DNS lab sample\\
%  3) Provide the third example (2d problem with species)\\
 In this section, we  present four demonstration cases for the application of TDB for the compression of streaming data: (i) Runge function; (ii) incompressible unsteady reactive flow; (iii) stochastic turbulent reactive flow, and (iv) three-dimensional turbulent channel flow.
%  we reduce and evolve a time dependent Rung function by our proposed TDB method to demonstrate the performance. After examining the error and adaptivity, we use our method to compress data from a three dimensional turbulent channel, and evolve it in time.  

 \subsection{Runge Function}
We first demonstrate the adaptive TDB compression technique with one dimensional modes (i.e., $d=p$) for a time dependent Runge function as follow:\\
 \begin{equation}\label{eq:Runge}
 f(x_1,x_2,x_3,t)=\frac{1}{a(t)^2+ x_1^2 + x_2^2 + x_3^2}, 
 \end{equation}
 %where $a(t)= 1 - 0.5 \exp(-\alpha(t-1)^2)$ and $\alpha=0.5$. For this choice of $a(t)$, the rank of the TDB decomposition must increase from $t=0$ to $t=1$ and then decrease  to ensure that  the low-rank approximation error remains below a specified value.  The spatial domain is the cube $[-\pi, \pi]^3$ discretized on a uniform grid of size $126^3$.   The time step $dt=0.005$ is used for evolving the TDB evolution equations. In this problem, we use the same number of modes in each direction, i.e., $r=r_1=r_2=r_3$ due to the isotropy of function $f$.  Figure \ref{Fig:SVdemo1} shows the reconstruction  error versus time for  fixed-rank TDB decompositions for $r=2$ and $r=3$. It also shows the error for adaptive TDB initiated with $r=3$. Figure \ref{Fig:SVdemo2} shows the singular values of the unfolded core tensor in the $x_1$-direction for the fixed-rank and adaptive-rank cases as well as the corresponding HOSVD singular values, which are obtained by taking  instantaneous SVD on the unfolded full-dimensional data. This shows that the fixed-rank and adaptive  TDB decompositions closely follow the HOSVD.   
 where $a(t)= 1 - 0.5 \exp(-\alpha(t-1)^2)$ and $\alpha=0.5$. For this choice of $a(t)$, the rank of the TDB decomposition must increase from $t=0$ to $t=1$ and then decrease  to ensure that  the low-rank approximation error remains below a specified value.  The spatial domain is the cube $[-\pi, \pi]^3$ discretized on a uniform grid of size $126^3$.   The time step $\Delta t=5 \times 10^{-3}$ is used for evolving the TDB evolution equations.

\begin{figure}
\centering
\subfigure[]{
\includegraphics[trim=2cm 0cm 2cm 0cm, clip=true ,width=0.3\textwidth]{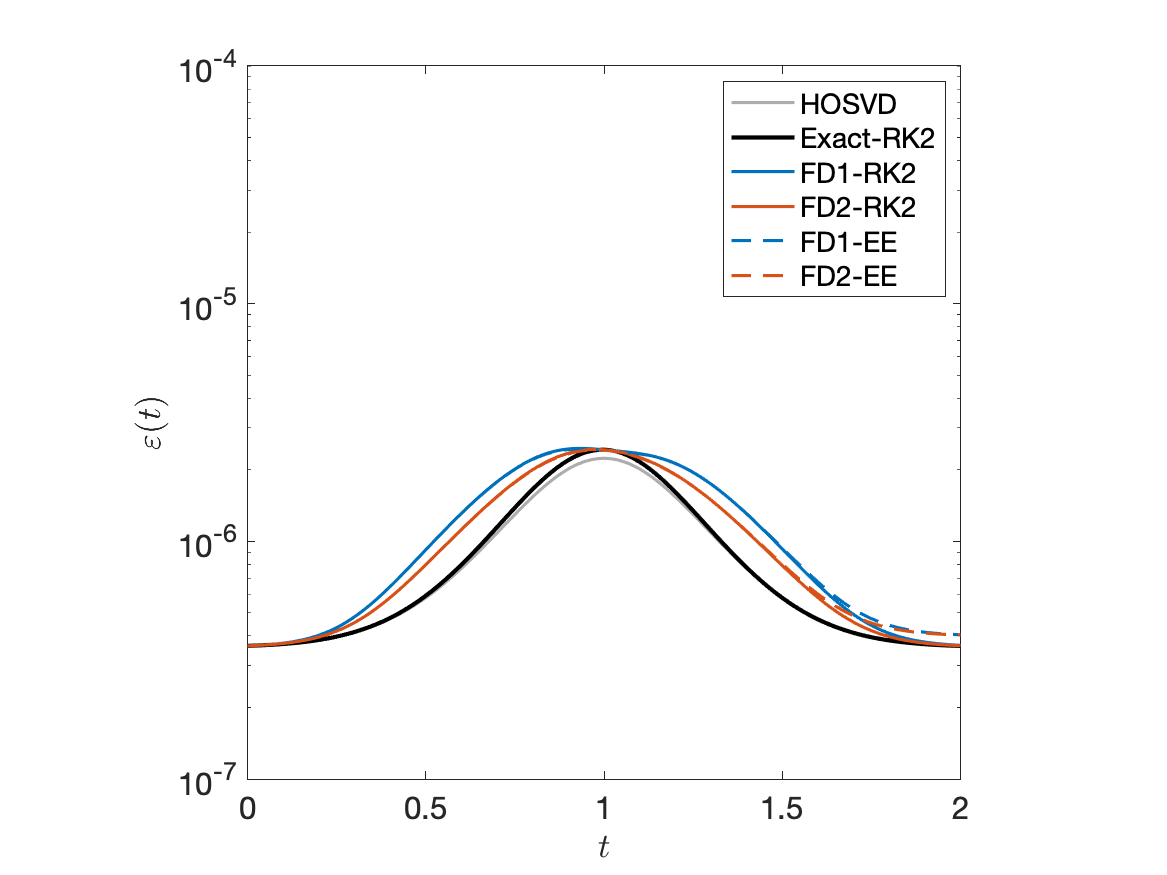}
\label{Fig:errorO}
}
\subfigure[]{
\includegraphics[trim=2cm 0cm 2cm 0cm, clip=true, width=0.3\textwidth]{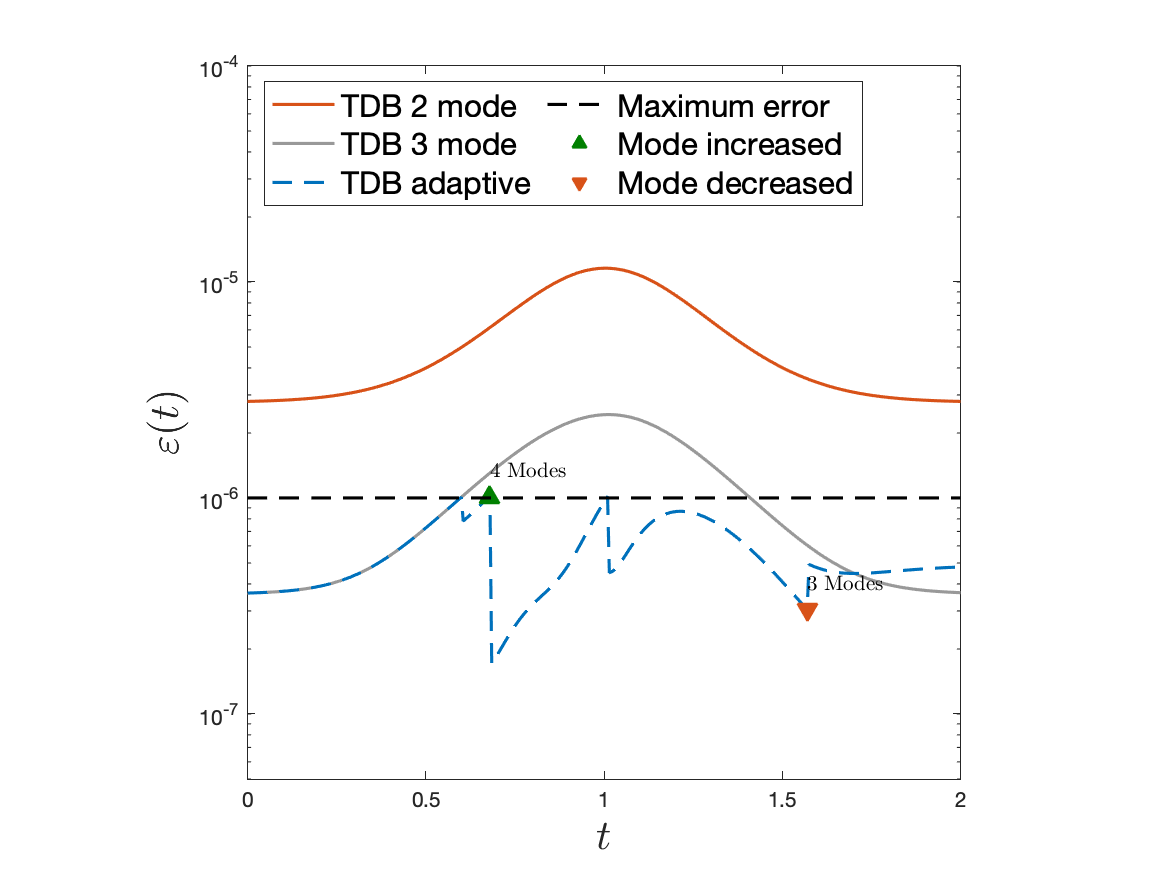}
\label{Fig:SVdemo1}
}
\subfigure[]{
\includegraphics[trim=2cm 0cm 2cm 0cm, clip=true ,width=0.3\textwidth]{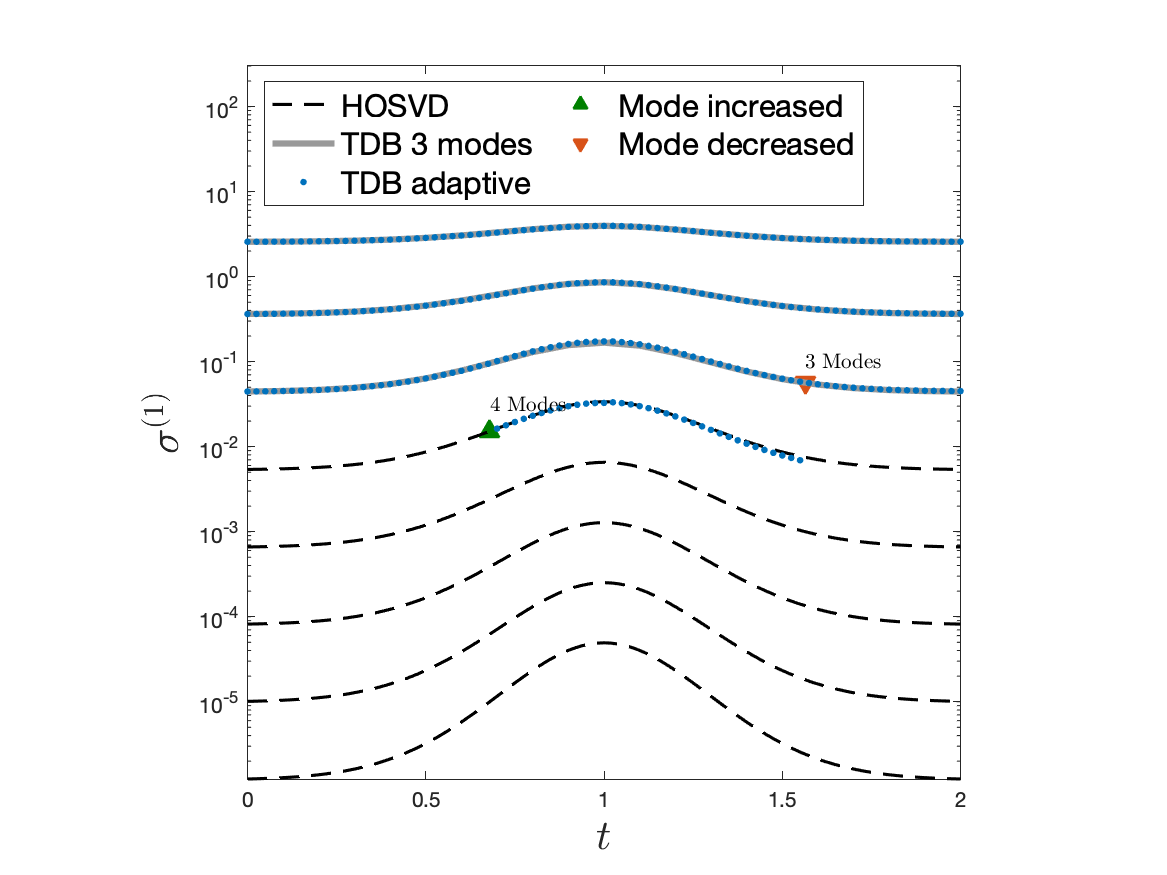}
\label{Fig:SVdemo2}
}

\caption{Runge function: (a) The effect of numerical and reduction error. (b) Comparing the effect of mode adjustment on error. (c) Singular values comparison.}
\label{Fig:SVdemo}

\end{figure}

 In this problem, we use the same number of modes in each direction, i.e., $r=r_1=r_2=r_3$ due to the isotropy of function $f$. First, we show the effect of both unresolved and numerical errors in figure \ref{Fig:errorO}. Here, HOSVD error is the weighted Frobenius norm of the difference between reconstructed data from HOSVD and the streaming data, which shows the unresolved error from reduction. The implemented first-order Euler (EE) and second-order Runge-Kutta (RK) add numerical error to the existing unresolved error. The second-order RK scheme has less error and we use this scheme to show adaptivity effect. This figure also shows the error difference between the exact data derivative with its first-order, and central second-order approximation. In order to study the adaptivity procedure for this analytical problem, we used the data derivative instead of finite difference approximations. Figure \ref{Fig:SVdemo1} shows the reconstruction  error versus time for fixed-rank TDB decompositions for $r=2$ and $r=3$ and adaptive TDB initiated with $r=3$. The error threshold and the captured energy percentage is set by practitioner to be $\varepsilon_{th}=10^{-6}$ and $\gamma_{th}=99.999\%$, respectively. It is evident that errors of the fixed-rank approximations $r=2$ and $r=3$ exceed the upper limit of the $10^{-6}$, while the adaptive TDB maintain the error below the upper limit by increasing the rank to $r=4$. The number of modes is later reduced to $r=3$ as the dimensionality of the problem decreases for the given $\gamma_{th}$. Figure \ref{Fig:SVdemo2} shows the singular values of the unfolded core tensor in the $x_1$-direction for the fixed-rank and adaptive-rank cases as well as the corresponding HOSVD singular values, which are obtained by taking instantaneous SVD on the unfolded full-dimensional data. This shows that the fixed-rank and adaptive TDB decompositions closely follow the HOSVD.\\

\subsection{Incompressible Turbulent Reactive Flow}

In the second demonstration, we apply the TDB compression to a turbulent reactive flow.  Turbulent reactive flows are multiscale and multivariate, whose high-fidelity numerical simulations result in massive datasets that with the current I/O restrictions of exascale simulations even storing the temporally resolved solution is becoming increasingly challenging let alone probing and analysis of the simulation data.  This is particularly the case when a large number of species is involved. An example is the direct numerical simulation (DNS) of turbulent combustion, in which the creation of an ignition kernel -- an intermittent phenomenon -- occurs on the order of 10 simulation time steps, while typically every 400th time step is stored to maintain the I/O overhead at a reasonable level \cite{BAB12}. 
 To demonstrate the application of TDB, we consider streaming data generated by a 2D advection-diffusion reaction problem:
\begin{equation}\label{ADR}
\frac{\partial \phi_i}{\partial t} + (u \cdot \nabla)  \phi_i = \nabla \cdot (\alpha_i \nabla  \phi_i) + Q_i^S
\end{equation}
where, $\phi_i$ is the concentration of $i$th reactant, $\alpha_i$ is the associated diffusion coefficient, $Q_i^S$ denotes the nonlinear source that determines whether $\phi_i$ is produced or consumed and $u$ is the velocity field of the flow.  The reaction mechanisms model the blood coagulation cascade in a Newtonian fluid \cite{LYTK15}. The simulation setup chosen in this work is identical to the case considered in reference \cite{RNB21}.  Equation (\ref{ADR})  is solved for $n_s=23$ species and the velocity field is obtained by solving incompressible Navier-Stokes equations at Re=1000 using spectral element discretization. Equation (\ref{ADR}) is advanced in time using fourth-order RK with $\Delta t=5 \times 10^{-4}$. The numerical solution of Eq. (\ref{ADR}) is cast as a streaming third-order tensor of size $N_1 \times N_2 \times N_3$, where $N_1=251$, $N_2= 76$ and $N_3= n_3=23$ are the number of species. 

\begin{figure}[t!]
\centering
\includegraphics[width=0.7\textwidth]{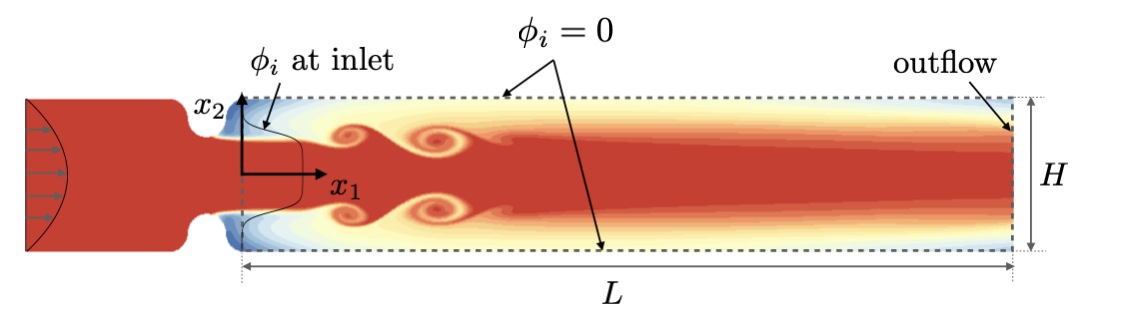}
\label{Fig:jet}
\caption{Incompressible turbulent reactive flow schematic.}
\end{figure}
We consider two different TDB compression schemes as follow: 
\begin{align} 
    &\mbox{TDB-1:} &\phi(x_1,x_2,\eta,t) = \sum_{i_3=1}^{r_3} \sum_{i_2=1}^{r_2} \sum_{i_1=1}^{r_1} \tn{T}_{i_1 i_2 i_3}(t)u^{(1)}_{i_1}(x_1,t)u^{(2)}_{i_2}(x_2,t)u^{(3)}_{i_3}(\eta,t)  \label{eq:1dreac} \\
    &\mbox{TDB-2 (DBO):}    &\phi(x_1,x_2,\eta,t) =  \sum_{i_2=1}^{r_2} \sum_{i_1=1}^{r_1} \tn{T}_{i_1 i_2}(t)u^{(1)}_{i_1}(x_1,x_2,t)u^{(2)}_{i_2}(\eta,t),  \label{eq:2dreac}
 \end{align}
where the inner product in the physical and composition space are as follow:
 \begin{align*}
    &\mbox{Physical space (TDB-1):} & \inner{u^{(1)}_{i_1}(x_1,t)}{u^{(1)}_{i'_1}(x_1,t)}_{x_1} \simeq {\bm{u}^{(1)}_{i_1}}^T \bm{W}^{(1)} \bm{u}^{(1)}_{i_1} \\ & & \inner{u^{(2)}_{i_2}(x_2,t)}{u^{(2)}_{i'_2}(x_2,t)}_{x_2} \simeq {\bm{u}^{(2)}_{i_2}}^T \bm{W}^{(2)} \bm{u}^{(2)}_{i_2}\\
    &\mbox{Physical space (TDB-2):} & \inner{u^{(1)}_{i_1}(\bm{x},t)}{u^{(1)}_{i'_1}(\bm{x},t)}_{\bm{x}} \simeq {\bm{u}^{(1)}_{i_1}}^T \bm{W}^{(1)} \bm{u}^{(1)}_{i_1}\\
    &\mbox{Composition space (TDB-1):}& \inner{u^{(3)}_{i_3}(\eta,t)}{u^{(3)}_{i'_3}(\eta,t)}_{\eta} \simeq {\bm{u}^{(3)}_{i_3}}^T  \bm{u}^{(3)}_{i_3}\\
     &\mbox{Composition space (TDB-2):}& \inner{u^{(2)}_{i_2}(\eta,t)}{u^{(2)}_{i'_2}(\eta,t)}_{\eta} \simeq {\bm{u}^{(2)}_{i_2}}^T  \bm{u}^{(2)}_{i_2}\\
 \end{align*}
 In TDB-1, $\bm{W}^{(1)}$ and $\bm{W}^{(2)}$ are quadrature weights obtained from spectral element discretizations of $x_1$ and $x_2$ directions, respectively. In TDB-2, $\bm{W}^{(1)}$ are the quadrature weights for 2D spectral element discretization i.e.,   both $x_1$ and $x_2$ directions. The inner product weight in the composition space for both schemes is the identity matrix. The above two schemes have different reduction errors and compression ratios. The main difference between the two decompositions is that in Eq. \ref{eq:1dreac}, the correlations between both spatial directions, i.e.,  $x_1$, $x_2$ and the \emph{composition} space, i.e.,  $\eta$  are extracted, whereas in TDB-2, the correlation between $x_1$ and $x_2$ directions are not extracted. Both schemes are adaptive: modes are added and removed to keep the reconstruction error below $\varepsilon_{th} = 10^{-5}$ and the captured details above the $\gamma_{th}=99.999\%$. TDB-2 was recently introduced in Ref. \cite{RNB21}, where it was referred to as \emph{dynamically bi-orthonormal} (DBO) decomposition since  the species  are decomposed to two sets of orthonormal modes in the spatial domain and the composition space.  In the DBO formulation presented in Ref. \cite{RNB21},  full-dimensional Eq. \ref{ADR} is not solved; instead closed-form evolution equations for all three components of TDB-2 are derived, i.e., ODEs for $\tn{T}_{i_1 i_2}(t)$ and $u^{(2)}_{i_2}(\eta,t)$ and PDEs for $u^{(1)}_{i_1}(x_1,x_2,t)$. We refer to DBO reduction in Ref. \cite{RNB21}, as model-driven analogue of the data-driven reduction technique presented in this paper, because in the DBO formulation in Ref. \cite{RNB21}  data generation is not required. In this section, we compare the model-driven DBO bases and the data-driven DBO bases that are the focus of this paper.

\begin{figure}[t!]
\centering
\subfigure[]{
\includegraphics[trim=2.5cm 0cm 2.5cm 0cm, clip=true, ,width=0.31\textwidth]{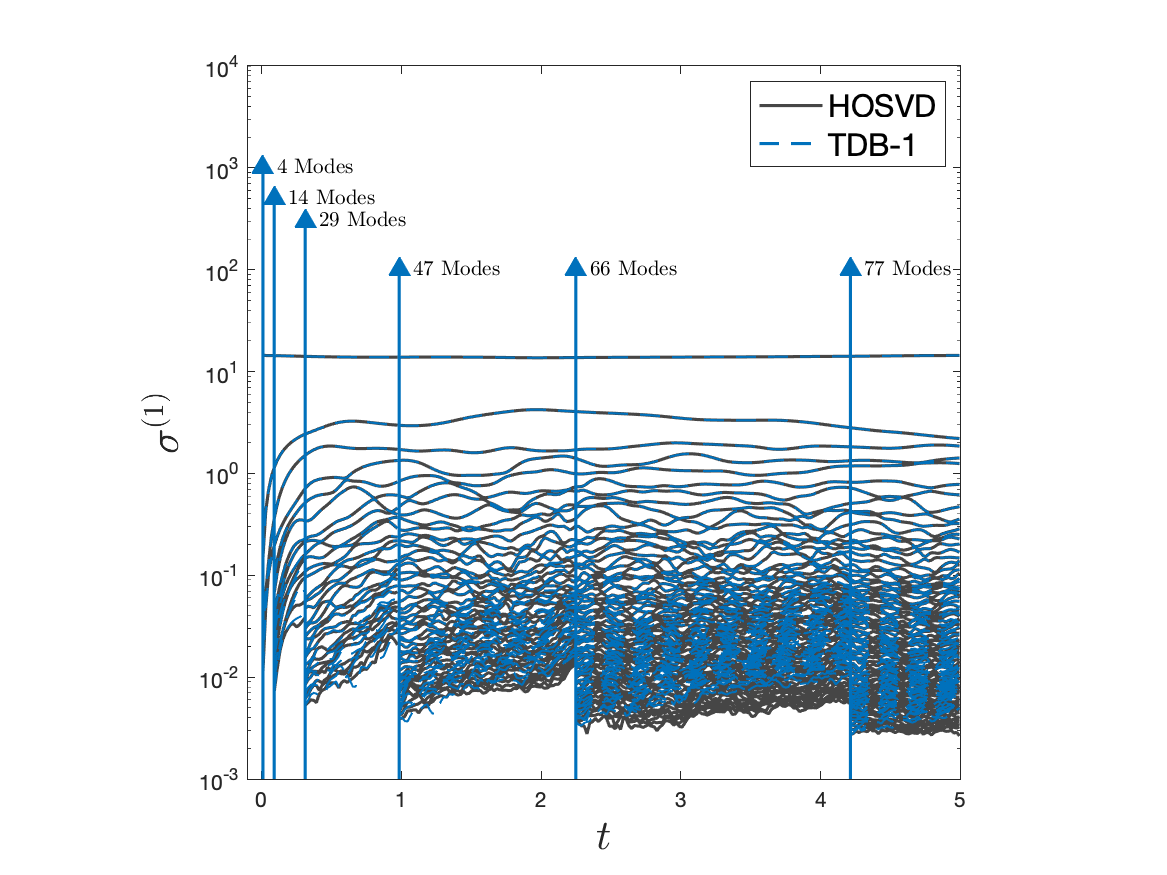}
\label{Fig:SVspX1}
}
\subfigure[]{
\includegraphics[trim=2.5cm 0cm 2.5cm 0cm, clip=true, ,width=0.31\textwidth]{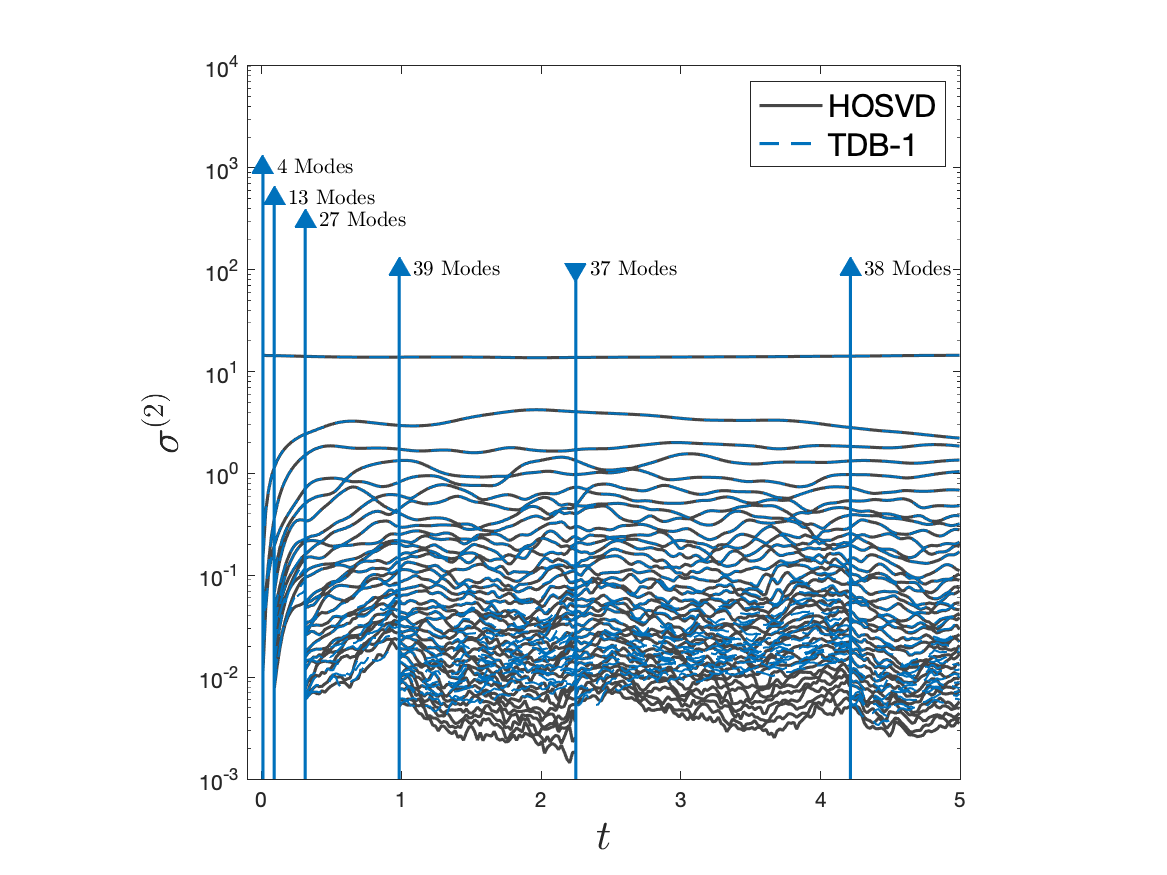}
\label{Fig:SVspX2}
}
\subfigure[]{
\includegraphics[trim=2.5cm 0cm 2.5cm 0cm, clip=true, ,width=0.31\textwidth]{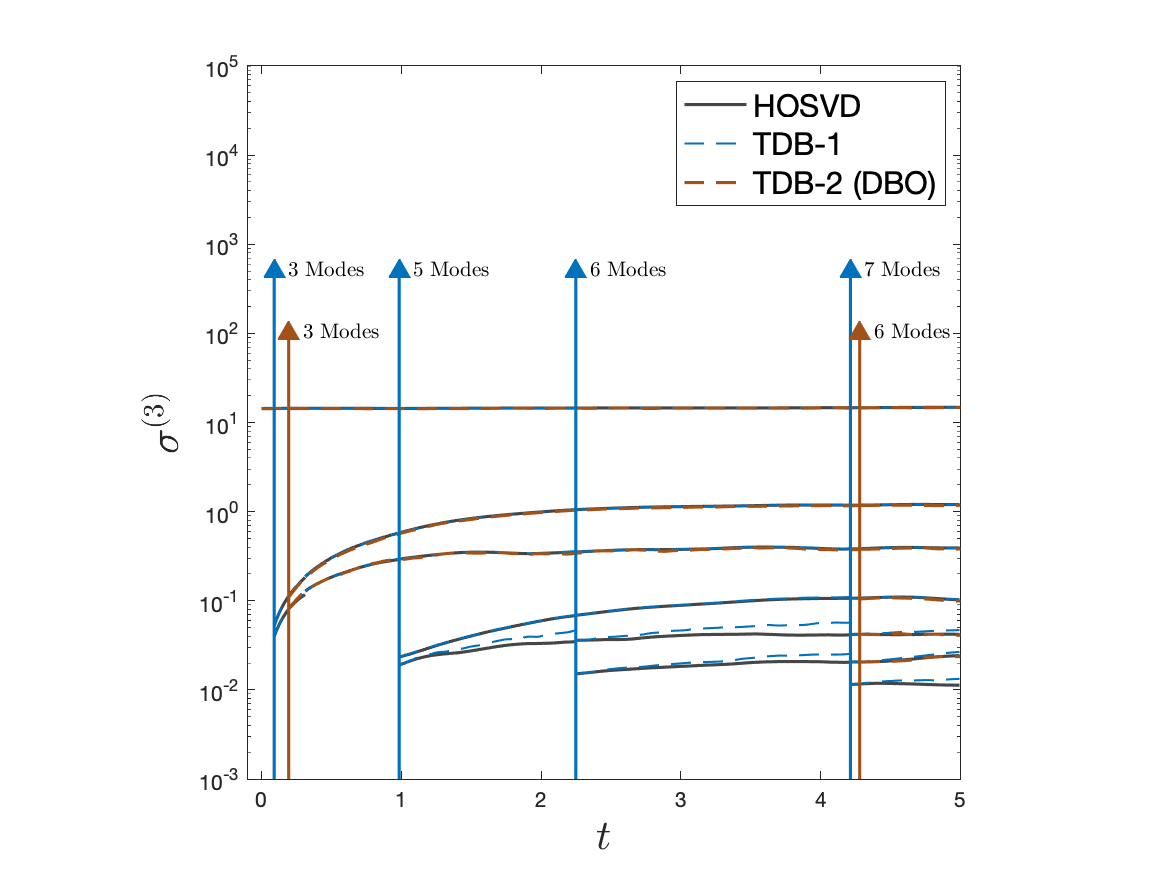}
\label{Fig:SVspX3}
}
\caption{Compression of the 2D turbulent reactive flow. Singular values of the unfolded tensor in directions: (a) $x_1$, (b) $x_2$, and (c) $\eta$.}
\label{Fig:SVsperror}
\end{figure}
\begin{figure}[t!]
\centering
\subfigure[]{
\includegraphics[trim=2.5cm 0cm 1.5cm 0cm, clip=true, width=0.35\textwidth]{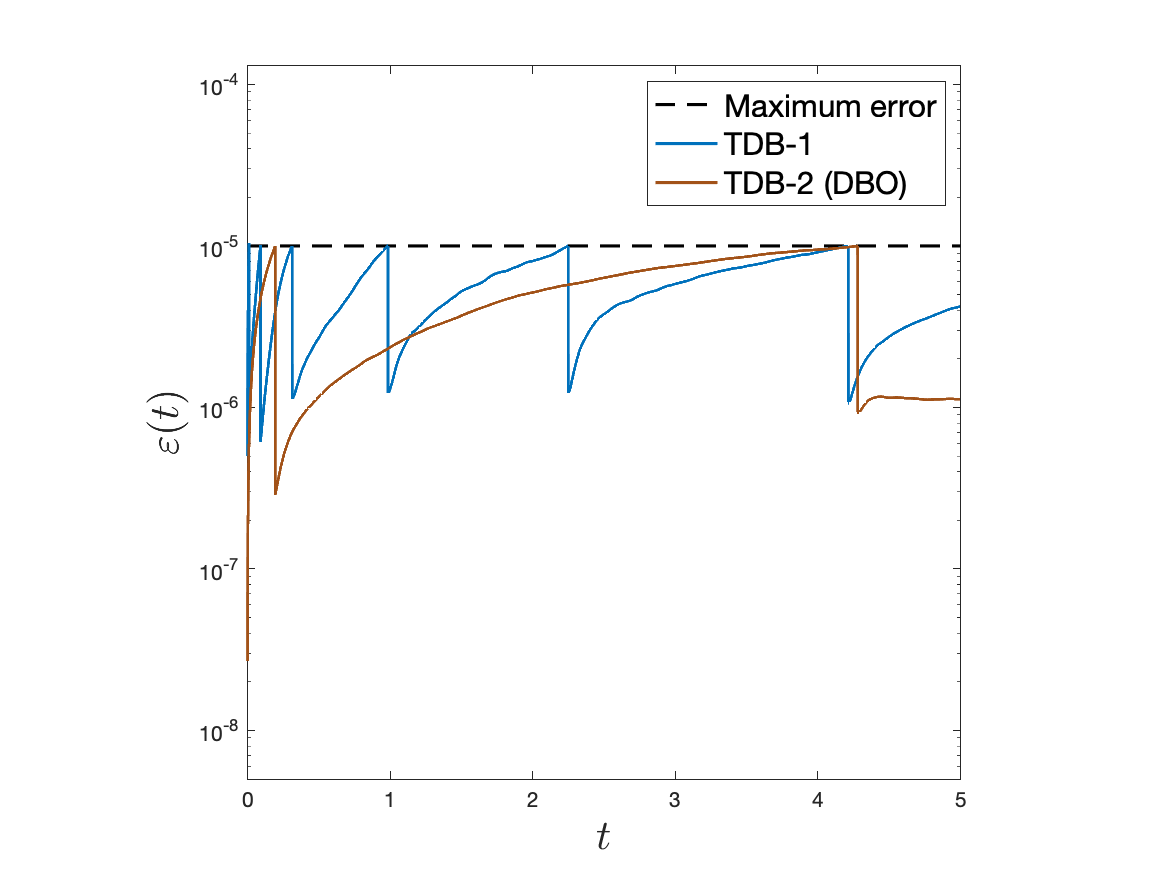}
\label{Fig:Sperror}
}
\subfigure[]{
\includegraphics[trim=2.5cm 0cm 1.5cm 0cm, clip=true, width=0.35\textwidth]{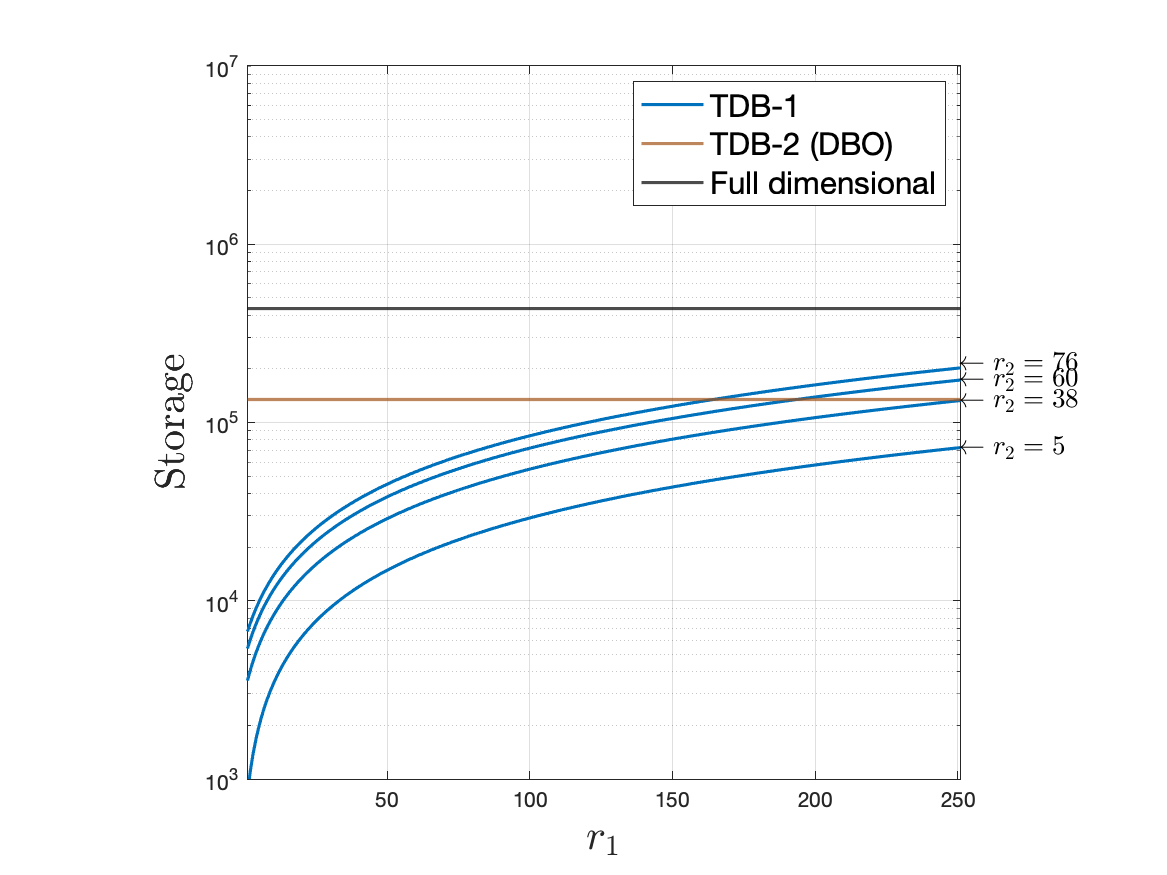}
\label{Fig:SpecComp}
}
\caption{Compression of turbulent reactive flow data using TDB-1 and TDB-2 (DBO) schemes: (a) error evolution, (b) storage cost.}
\label{Fig:Sperror&comp}
\end{figure}

TDB-2 shows a slower error growth than TDB-1. To investigate this, in Figures \ref{Fig:SVspX1}-\ref{Fig:SVspX3}, we show   the instantaneous  singular values of the unfolded tensor in $x_1$, $x_2$ and $\eta$ directions for TDB-1 and TDB-2 as well as  the instantaneous singular values obtained by performing HOSVD. Since in the TDB-2 decomposition, unfolding in $x_1$ and $x_2$
directions are not formed, in Figures \ref{Fig:SVspX1}-\ref{Fig:SVspX2} only the singular values of  TDB-1 and HOSVD can be shown.  It is clear that dominant singular values are captured accurately by both schemes.  However, the effect of unresolved modes introduces a memory error. This error  is driven by the energy of the unresolved modes and it affects  the lower-energy modes more intensely than the higher-energy modes.  If this error is left uncontrolled, it will eventually  contaminate the higher-energy modes. The reconstruction error evolution for both schemes is shown in Figure  \ref{Fig:Sperror}. It is evident that the error grows at a much faster rate in TDB-1 compared to TDB-2. This can be explained by observing that the dimensionality in the spatial domain is often much higher than the dimensionality in the composition space. This means that in TDB-1, $\sigma^{(3)}(t)$ has a much faster decay rate compared to  $\sigma^{(1)}(t)$ and $\sigma^{(2)}(t)$ as can be clearly seen in Figures \ref{Fig:SVspX1}-\ref{Fig:SVspX3}. In TDB-2, there is  no reduction error  in the $x_1$ or $x_2$ directions, and moreover, the turbulent reactive flow shows a very low-dimensional dynamics in the composition space ($\eta$), as a result the overall energy of the unresolved modes is much smaller, which in turn leads to a slower error growth rate. This also means that in an adaptive mode, one has to do  calculate HOSVD  more often in TDB-1 than in TDB-2.    \\

\begin{figure}[t!]
\centering
\includegraphics[width=0.95\textwidth]{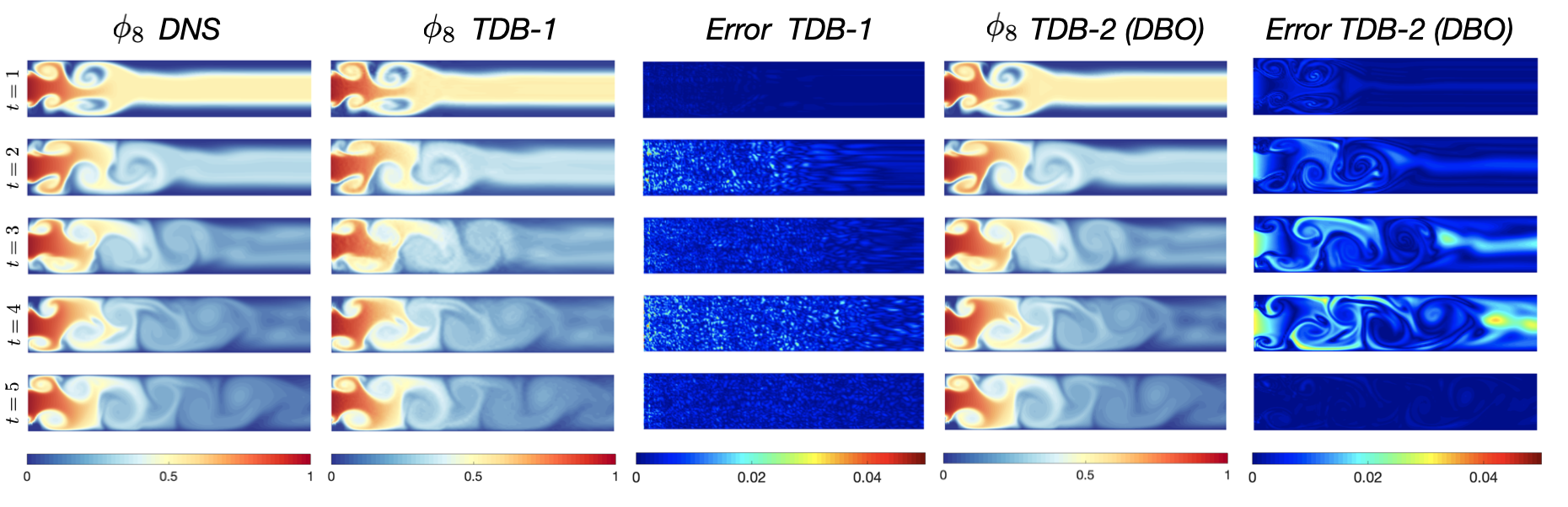}
\caption{Unsteady reactive flow: Comparison between DNS and TDB reconstructed species concentration.}
\label{Fig:SpeciesDataVSTDB}
\end{figure}

In Figure \ref{Fig:SpeciesDataVSTDB}, we compare the reconstructed data for the eighth species at five time instants. The reconstructed data are in good agreement with DNS snapshots, however, the form of error is different. The error is computed as the absolute value of the difference between the DNS and reconstructed TDB data. The resulting error from TDB-1 appears to have much finer structure than that of TDB-2. That is because in TDB-1, the unresolved subspaces in each of the $x_1$ and $x_2$ dimensions have very fine structures and they dominate the error, whereas in TDB-2 the error is due to the unresolved subspace in the $\eta$ direction.  The TDB-1 error at $t=1,2,3,4$ is less than TDB-2, mainly because of frequent HOSVD reinitialization in the adaptivity process. However, the maximum error of TDB-1 is larger than that of TDB-2. For better comparison, the error color bar is set from 0 to 0.5, while the maximum errors for  TDB-1 and TDB-2 are 0.14 and 0.4, respectively.\\

TDB-1 and TDB-2 have different storage costs for the same reconstruction errors.  In TDB-2, slower error growth rate comes  at the cost of storing $r$ two-dimensional bases with the storage cost of $r N_1 N_2$ as opposed to TDB-1, where the storage cost of the spatial bases is $r_1 N_1 + r_2 N_2$. The overall storage cost of TDB-1 is: $S_{TDB-1} = r_1 N_1 + r_2 N_2 + r_3 n_s + r_1 r_2 r_3$ and the storage cost of TDB-2 is: $S_{TDB-2}= r N_1 N_2 + r n_s + r^2$. The comparison of the storage cost of TDB-1 and TDB-2 for the same reconstruction error depends on the values of $r_1$ and $r_2$.   These two storage costs for the same  number of modes in the composition space, i.e., $r=r_3=7$ are shown in Figure \ref{Fig:SpecComp}. If $r_1$ and $r_2$ are large enough, the storage cost of TDB-1  exceeds that of TDB-2.
In our study, the weighted compression ratio are $\overline{CR}_{TDB-1}=15.62$ and $\overline{CR}_{TDB-2}=6.6$ where the total size of data $35GB$ compressed to $2.2GB$ and $5.3GB$ by TDB-1 and TDB-2, respectively.\\

\begin{figure}[t!]
\centering
\includegraphics[width=0.95\textwidth]{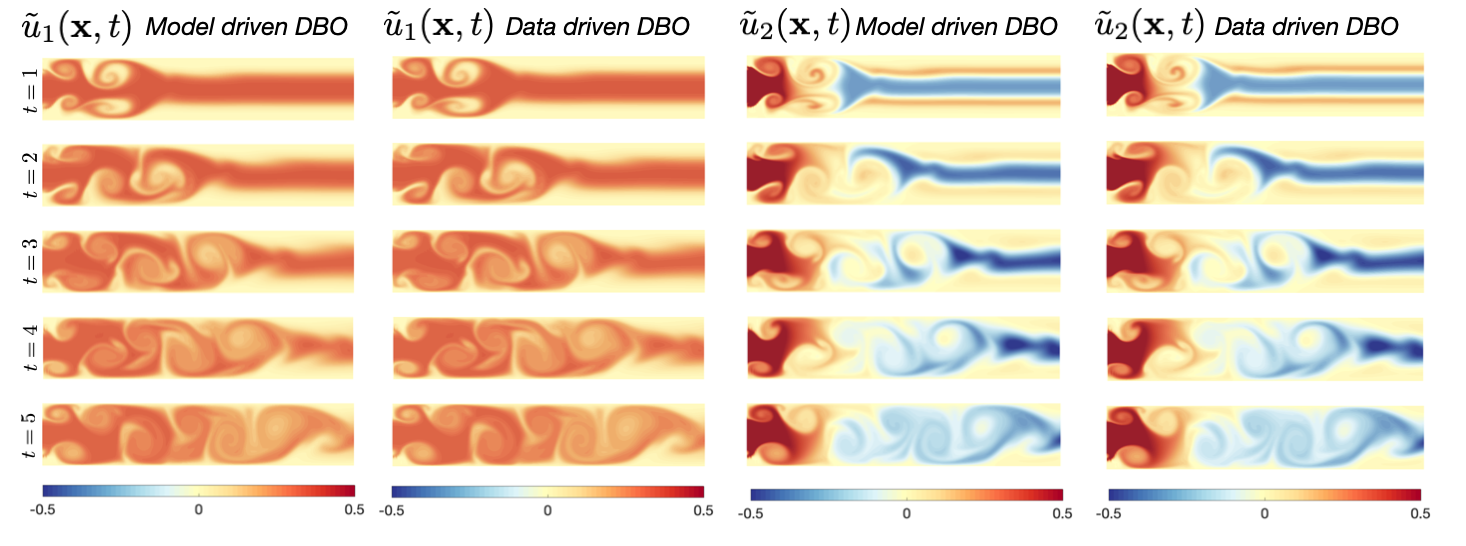}
\caption{Turbulent reactive flow: Comparison between the first two dominant modes of the model-driven and data-driven DBO.}
\label{Fig:DBOmodes}
\end{figure}

 In the DBO formulation presented in Ref. \cite{RNB21}, a coupled set of PDEs for the 2D bases and ODEs for the low-rank matrices are solved without using any data. The goal of the model-driven DBO \cite{RNB21} is to reduce the computational cost and memory requirement of solving species transport equations as well as reducing the I/O load.  In this work, our goal is to compress the streaming data generated by the full-dimensional model.  In Figure \ref{Fig:DBOmodes}, the first and second most dominant modes of model-driven and data-driven DBOs for different instances of times are shown. It is clear that the bases for both model-driven DBO and data-driven DBO are nearly identical.

\subsection{Stochastic Turbulent Reactive Flow}
\begin{figure}[t!]
\centering
\subfigure[]{
\includegraphics[trim=0cm 0cm 1cm 0cm, clip=true, width=0.35\textwidth]{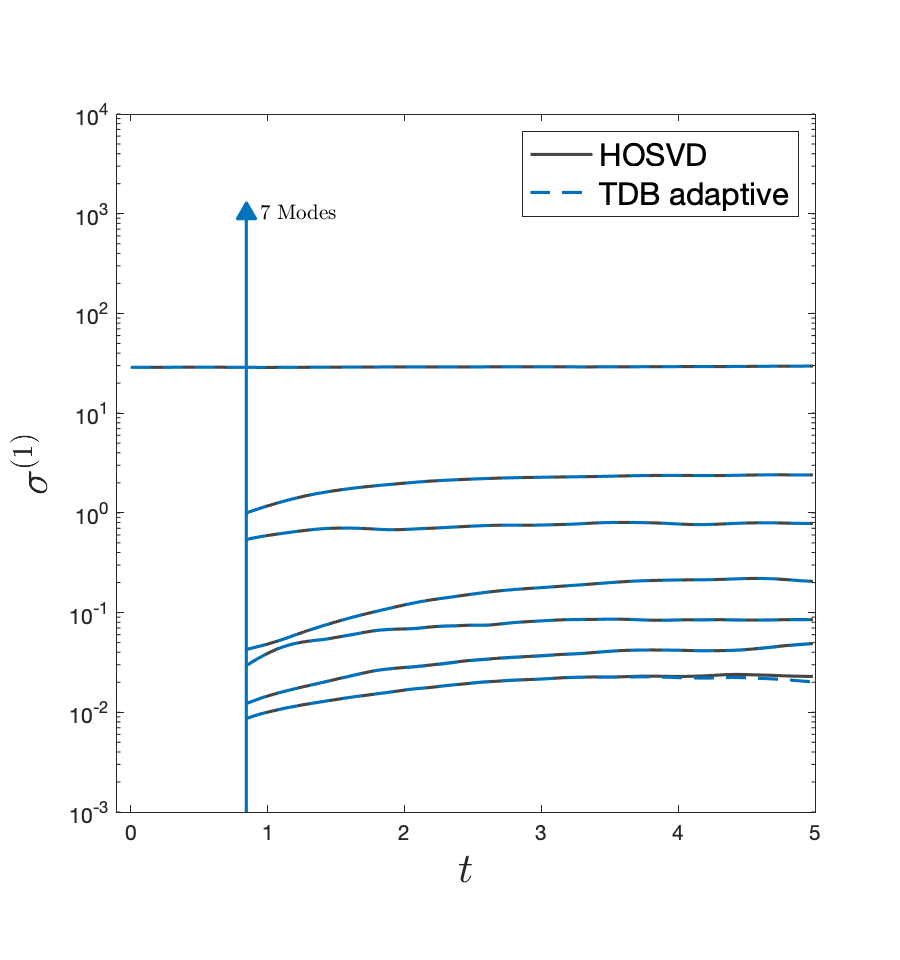}
\label{Fig:UQx1}
}
\subfigure[]{
\includegraphics[trim=0cm 0cm 1cm 0cm, clip=true, width=0.35\textwidth]{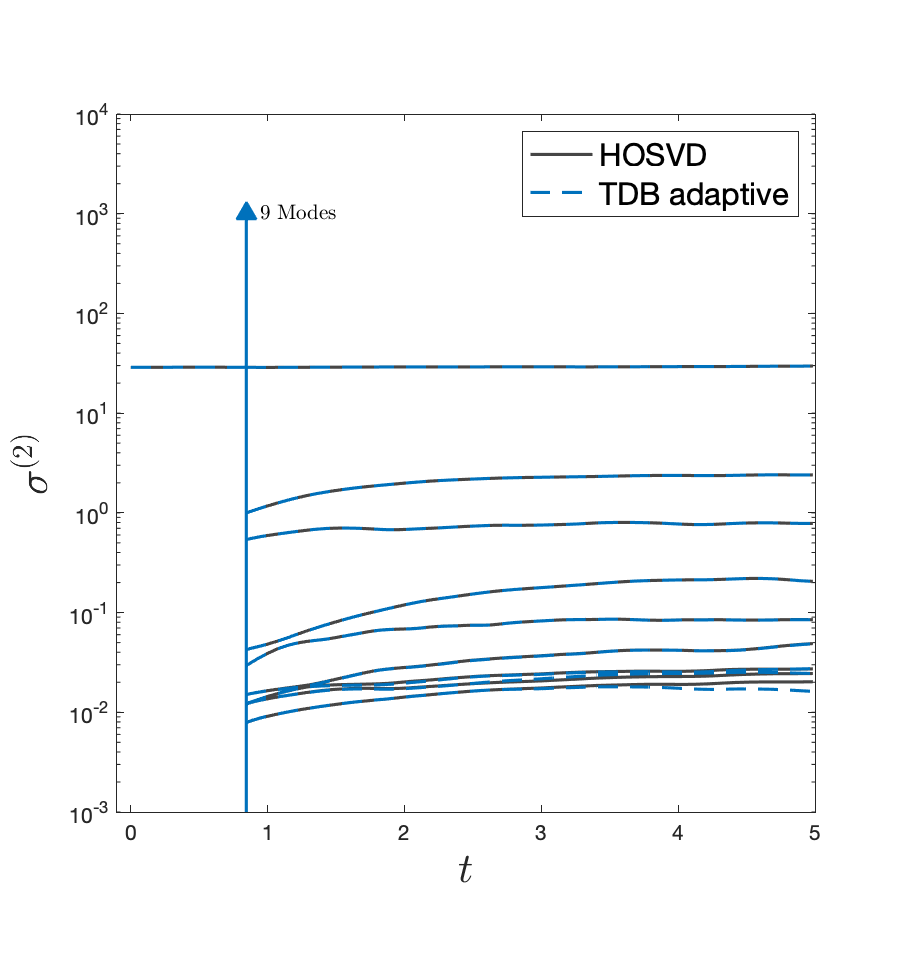}
\label{Fig:UQx2}
}
\subfigure[]{
\includegraphics[trim=0cm 0cm 1cm 0cm, clip=true, width=0.35\textwidth]{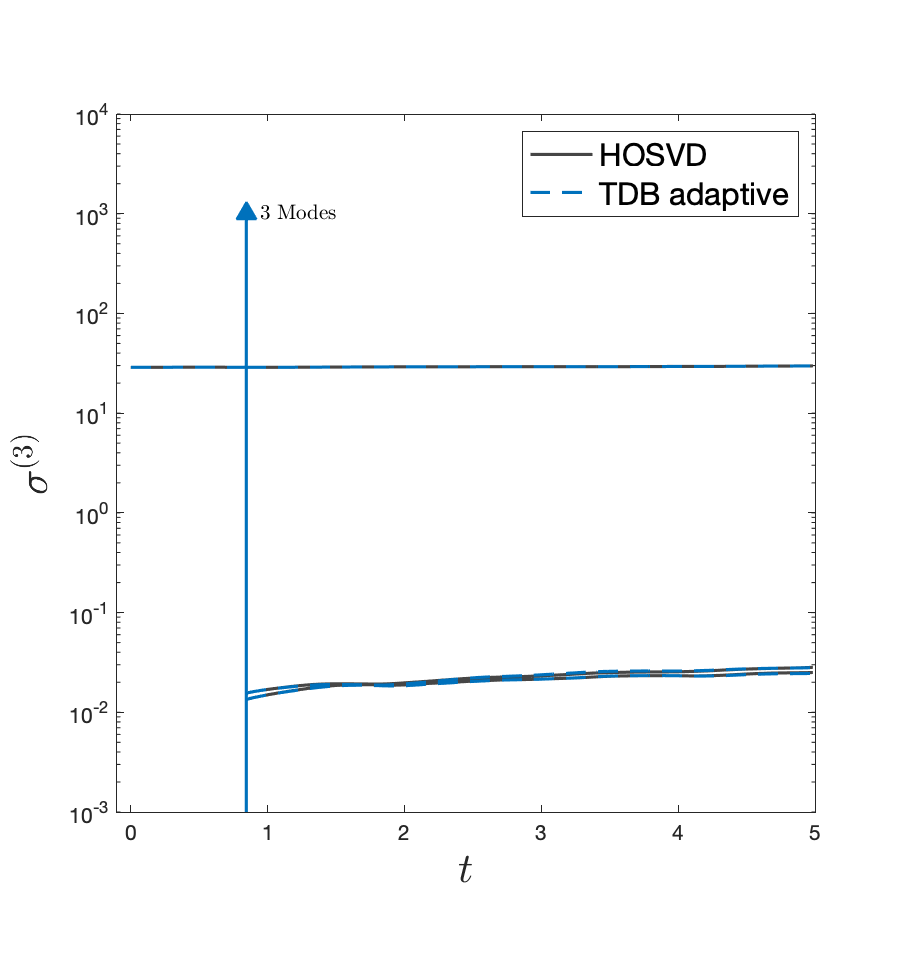}
\label{Fig:UQx3}
}
\subfigure[]{
\includegraphics[trim=0cm 0cm 1cm 0cm, clip=true, ,width=0.35\textwidth]{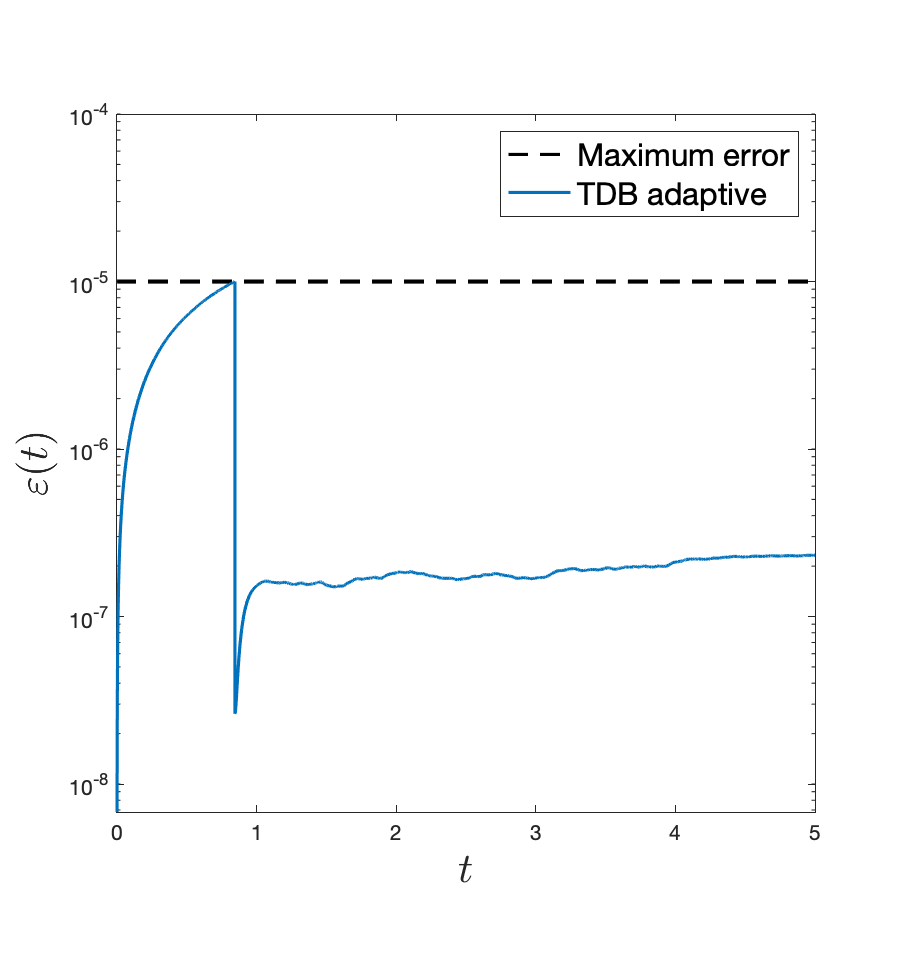}
\label{Fig:UQerror}
}
\caption{Incompressible turbulent reactive flow uncertainty quantification. Singular values of the unfolded tensor in: (a) $x_1$, (b) $x_2$, (c) $\eta$, and (d) error evolution.}
\label{Fig:UQresults}
\end{figure}
Quantifying uncertainties of transport and chemistry model parameters  has major implications for the field of chemically reactive flows.    An uncertainty quantification  (UQ) analysis can  significantly reduce the experimental costs by effectively allocating limited resources on reducing the uncertainty of key parameters, and inform mechanism reduction  by determining the least important parameters or detecting reaction pathways that    are unimportant and can be eliminated \cite{BOR15,LCR19,LLC20}.  Any sampling-based technique, e.g., Monte Carlo or probabilistic collocation methods (PCM) for UQ analysis of this problem can generate very large datasets. In this demonstration, we show how TDB can be utilized to extract correlation between different random samples in addition to the multidimensional  correlations presented in the previous example, to reduce the storage cost of the data generated.  
% Analyzing the effect of random parameters has a critical role in engineering problems which  requires a massive storage capacity. However, because of the high correlation between parameters, the generated data are highly compressible. 
In this section, we study the compression of resulting data from incompressible reactive flow with random diffusion coefficients. We consider the problem of  uncertainty diffusion coefficients for two species   ($\alpha_1$ and $\alpha_4$) in Eq. \ref{ADR}. We assume both of these two coefficients are independent  random variables as in the following:
\begin{equation*}
   \pmb{\xi}= [\xi_1,\xi_2]=(1 + 0.05 \omega) [\alpha_1,\alpha_4],
\end{equation*}
where $\omega$ is a uniform random variable $\sim \mathscr{U} [ -1,1 ]$. The rest of the problem setup remains identical to the problem considered in \S 3.2. We consider a collocation grid of size $s=4 \times 4$ in the random space, which requires 16 forward DNS simulations. We consider the following TDB compression scheme:
\begin{equation} 
    \phi(\bm{x},\eta,\pmb{\xi},t) = \sum_{i_3=1}^{r_3} \sum_{i_2=1}^{r_2} \sum_{i_1=1}^{r_1} \tn{T}_{i_1 i_2 i_3}(t)u^{(1)}_{i_1}(x_1,x_2,t)u^{(2)}_{i_2}(\eta,t) u^{(3)}_{i_3}(\xi_1,\xi_2,t).  \label{eq:UQ}
\end{equation}
The rationale for choosing a 2D TDB for the random space rather than two 1D TDBs is that the cost of storing 2D random bases is negligible. On the other hand,  choosing two 1D random TDBs would have increased the order of the core tensor from 3 to 4. Another valid choice for TDB is  to combine the composition space and the random direction into a 3D space. However, the interpretability of 1D TDB in the composition space  is quite appealing. The $u^{(2)}_{i_2}(\eta,t)$'s represent a time-dependent subspace in the composition space.
In which the inner product in the random space is defined as follow:\\
\begin{equation*}
    \inner{u^{(3)}_{i_3}(\xi_1,\xi_2,t)}{u^{(3)}_{i'_3}(\xi_1,\xi_2,t)}_{\pmb{\xi}}=\mathbb{E}[u^{(3)}_{i_3}(\xi_1,\xi_2,t) \: u^{(3)}_{i'_3}(\xi_1,\xi_2,t)] \simeq {\bm{u}^{(3)}_{i_3}}^T \bm{W}^{(3)} \bm{u}^{(3)}_{i_3},
\end{equation*}
where $\bm{W}^{(3)}$ is the probabilistic collocation weight. Each random TDB in the discrete form is represented by a vector of size $s$, i.e., $\bm{u}_i^{(3)} \in \mathbb{R}^{s\times 1}$.  Different sampling schemes might be used here. For example, one can use Monte Carlo samples. In that case, the inner product weight would be a diagonal matrix with all diagonal  entries equal to $1/s$. \\

Since there is a high degree of correlation between random samples of species fields, the TDB compression can achieve the high compression ratio $\overline{CR}=42.75$ compared to the previous cases and compress the size of data from $561.5GB$ to $13.1GB$. To examine this, we show the resulting singular values of the unfolded core tensor and compare them with HOSVD singular values in Figures \ref{Fig:UQx1}-\ref{Fig:UQx3}. Based on these figures, we make the following observations: (i) the dominant singular values are captured accurately; (ii) similar to TDB-2 the singular values decay rate is fast for this compression scheme compared to TDB-1; (iii) the problem dimension does not change after $t=1$, and therefore, the number of modes remains the same; (iv) the random space has the lowest dimensionality, which means the generated data are highly correlated with respect to the random diffusion coefficients. Figure \ref{Fig:UQerror} shows the reconstruction error evolution in which it exceeds the maximum limit around $t=1$ due to the increase in dimensionality. Using the reinitialization and mode adjustment in each direction with respect to the defined $\gamma_{th}=99.999\%$ the error decreases. Since after $t=1$ the dimensionality does not change, the error remains the same.

\subsection{Three-dimensional Turbulent Channel Flow}
In the last demonstration case, we use TDB to compress the data obtained by the direct numerical simulation (DNS) of turbulent channel flow. The data are generated by the finite difference solver in Ref.  \cite{vuorinen2016dnslab}. The Reynolds number based on the friction velocity is $Re_{\tau}=180$. The  length, width and height of the channel are $\pi$, $2\pi$, and $2$, respectively. The number of grid points in all dimensions is $150$ with uniform distribution in streamwise and spanwise directions. The grid is clustered near the channel wall in the wall-normal direction. 
\begin{figure}[t!]
\centering
\subfigure[]{
\includegraphics[trim=2.5cm 0cm 2.5cm 0cm, clip=true, ,width=0.43\textwidth]{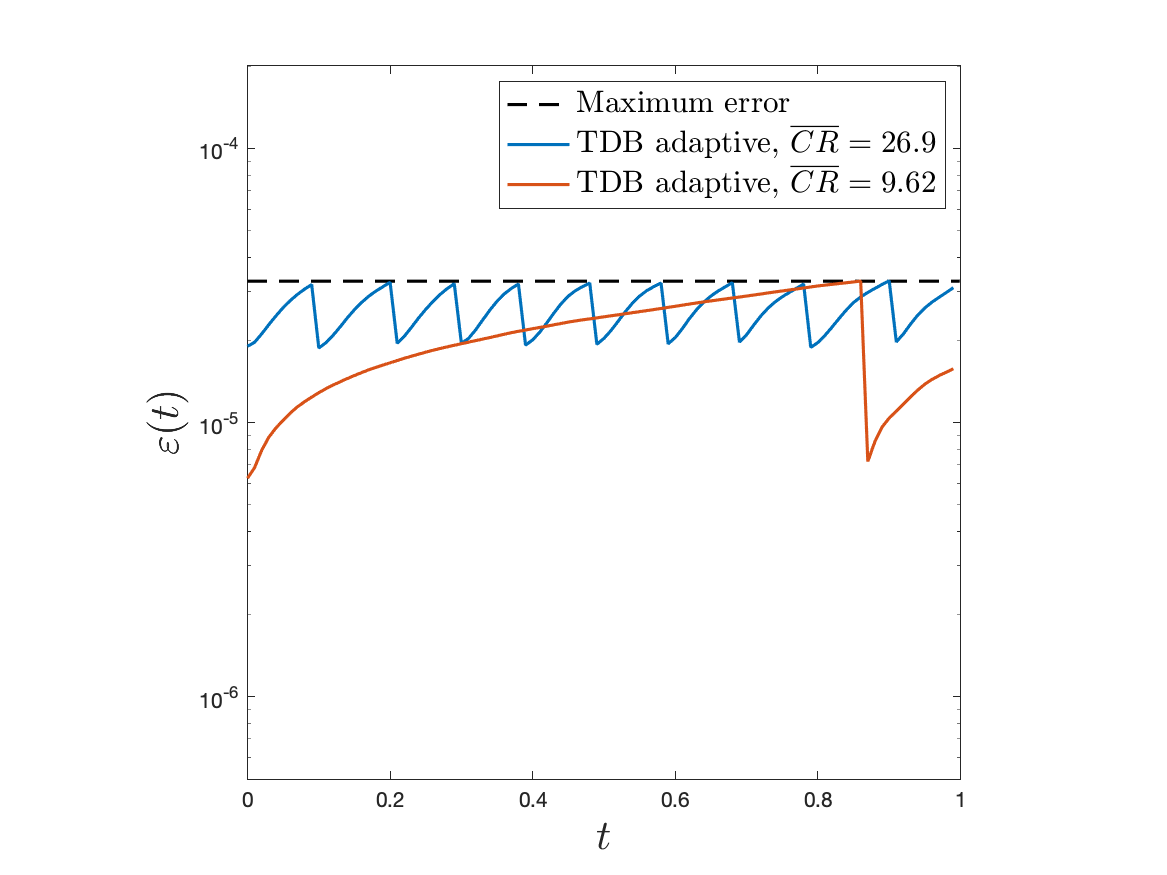}
\label{Fig:Channelerror}
}
\subfigure[]{
\includegraphics[trim=2.5cm 0cm 2.5cm 0cm, clip=true, width=0.43\textwidth]{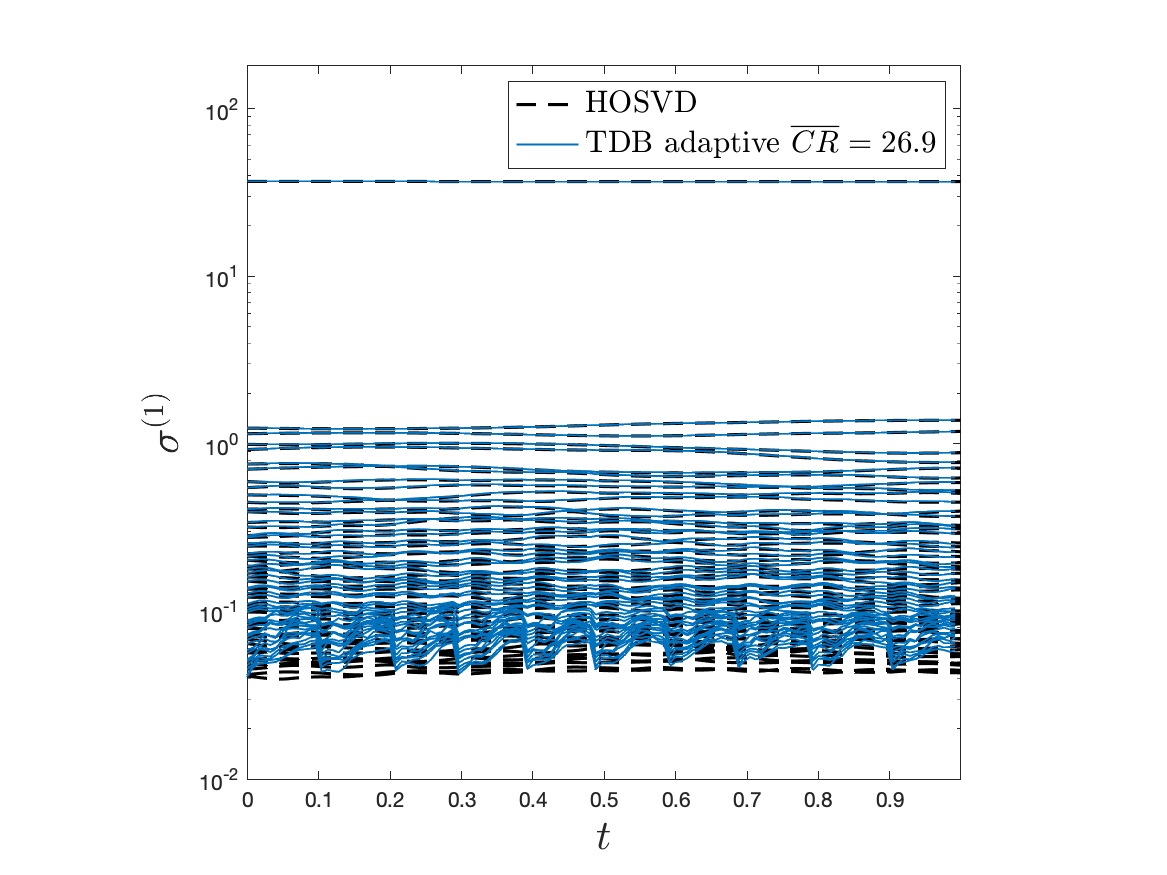}
\label{Fig:ChannelSV}
}
\caption{Turbulent channel flow: (a) Error evolution. (b) Singular values of the unfolded tensor in $x_1$ direction.}
\label{Fig:Channelerror&SV}
\end{figure}
We  apply TDB to the streamwise velocity component as compression of other field variables result in the same qualitative observations. We consider one-dimensional TDBs as follows:
\begin{equation} 
     v(x_1,x_2,x_3,t) = \sum_{i_3=1}^{r_3} \sum_{i_2=1}^{r_2} \sum_{i_1=1}^{r_1} \tn{T}_{i_1 i_2 i_3}(t)u^{(1)}_{i_1}(x_1,t)u^{(2)}_{i_2}(x_2,t)u^{(3)}_{i_3}(x_3,t).  \label{eq:TDB_DNS}
 \end{equation}
The TDB equations are evolved with the time step of $\Delta t = 10^{-3}$, which is the same as the $\Delta t$  used in the DNS time integration.  We consider two cases with different error thresholds. Case-I with the $\gamma_{th}=99.9999\%$  and Case-II with the $\gamma_{th}=99.999\%$. For both cases the error threshold is set to be $\varepsilon_{th}=3.3 \times 10^{-5}$. In Case-I, the initial value of the multirank is  $(r_1,r_2, r_3) = (68,81,50)$  and in  Case-II the initial  multirank is $ (r_1,r_2, r_3) = (54,57,36)$. The error evolution for both case are shown  in Figure  \ref{Fig:Channelerror}.  The  compression ratios for Case-I and Case-II are $\overline{CR}=9.62$ and $\overline{CR}=26.90$, respectively.  For this problem, performing HOSVD at different time instants result in the same multirank  and that is because the flow is statistically steady state, and therefore, the rank of the systems does not change in time.  However, in the TDB formulation, the error still grows because of the effect of the unresolved subspace. This error is controlled by the adaptive scheme. The error evolution for both cases are shown in Figure \ref{Fig:Channelerror}.  In Case-II, since the unresolved space has larger energy, the error grows with faster rate compared to Case-I. As a result, in Case-II, 9 HOSVD initializations are performed as opposed to one initialization in Case-I. 
% When the error exceeds the upper bound of the error buffer zone, the adaptive procedure  reinitialize the basis with almost the same number of modes. Since, the error does not drop bellow the error buffer zone, there is no rank reduction. The error increment is due to the high dimensionality and unresolved modes. In this case, the resulting small singular values have similar energy levels, and error in their evolution can shift them to higher level of energy.
Figure \ref{Fig:ChannelSV} shows the singular values of the  unfolded TDB core tensor along the $x_1$ direction against that of the full-dimensional data obtained by  HOSVD. There is a large gap between the first singular value and the rest of the singular values. We also observe that the low-energy modes values are more contaminated by the error induced by the unresolved subspace. 

\begin{figure}[t!]
\centering
\subfigure[]{
\includegraphics[trim=1cm 0cm 1cm 0cm, clip=true, width=0.45\textwidth]{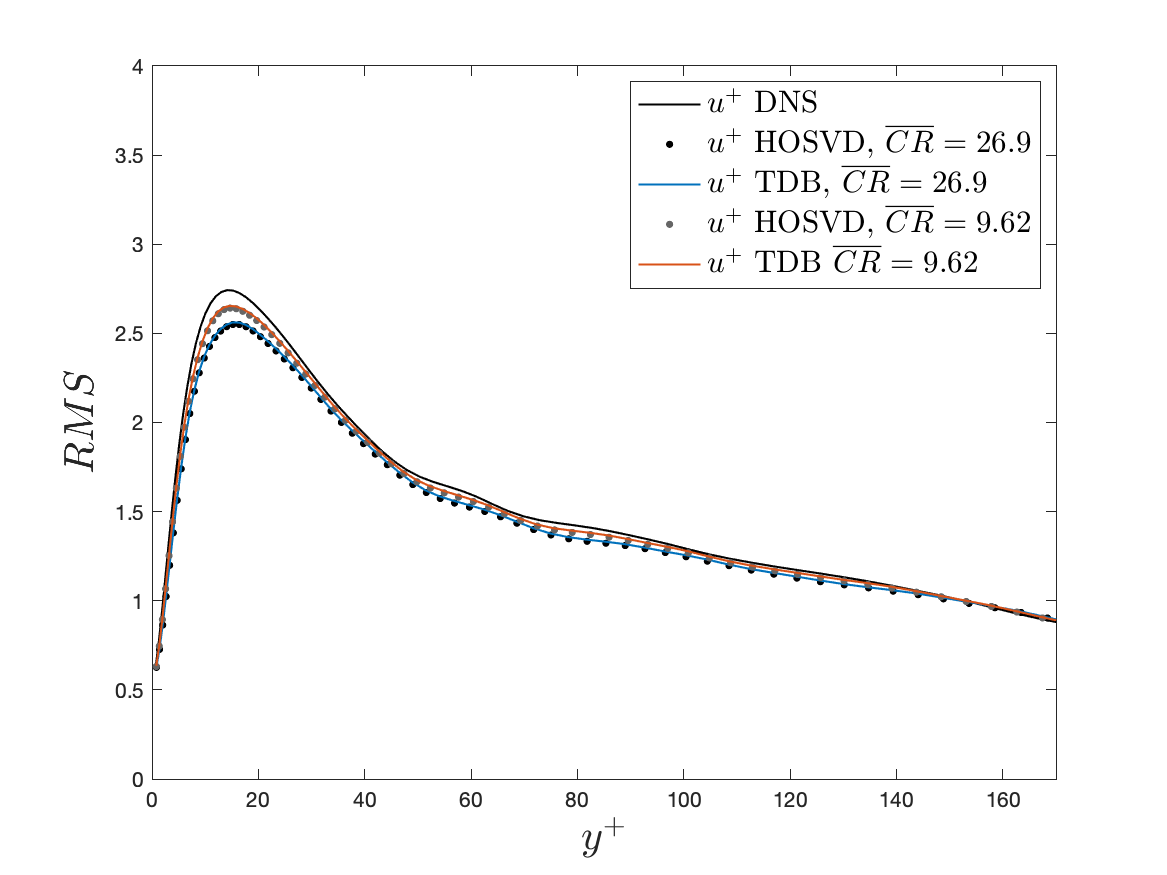}
\label{Fig:RMS}
}
\subfigure[]{
\includegraphics[trim=1cm 0cm 1cm 0cm, clip=true, width=0.45\textwidth]{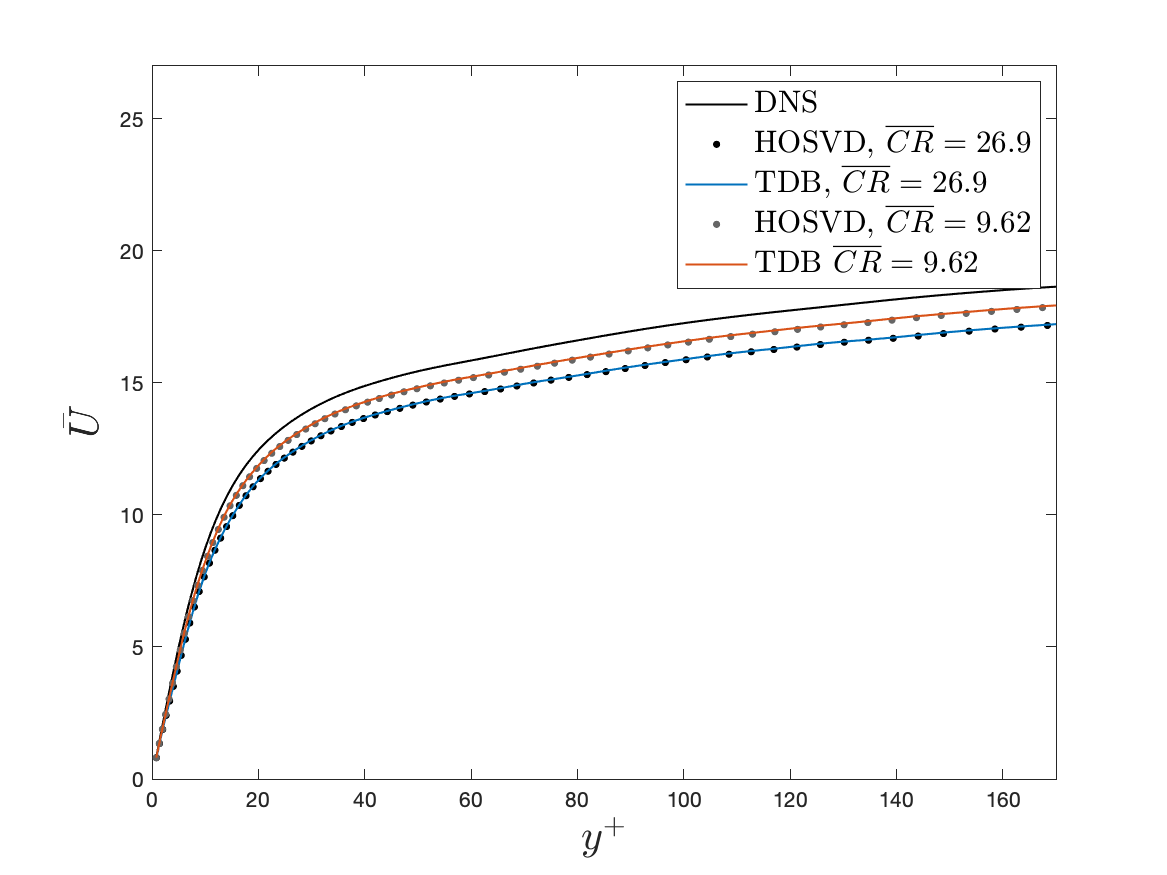}
\label{Fig:Mean}
}
\subfigure[DNS]{
\includegraphics[trim=0.5cm 0cm 1cm 0cm, clip=true, width=0.3\textwidth]{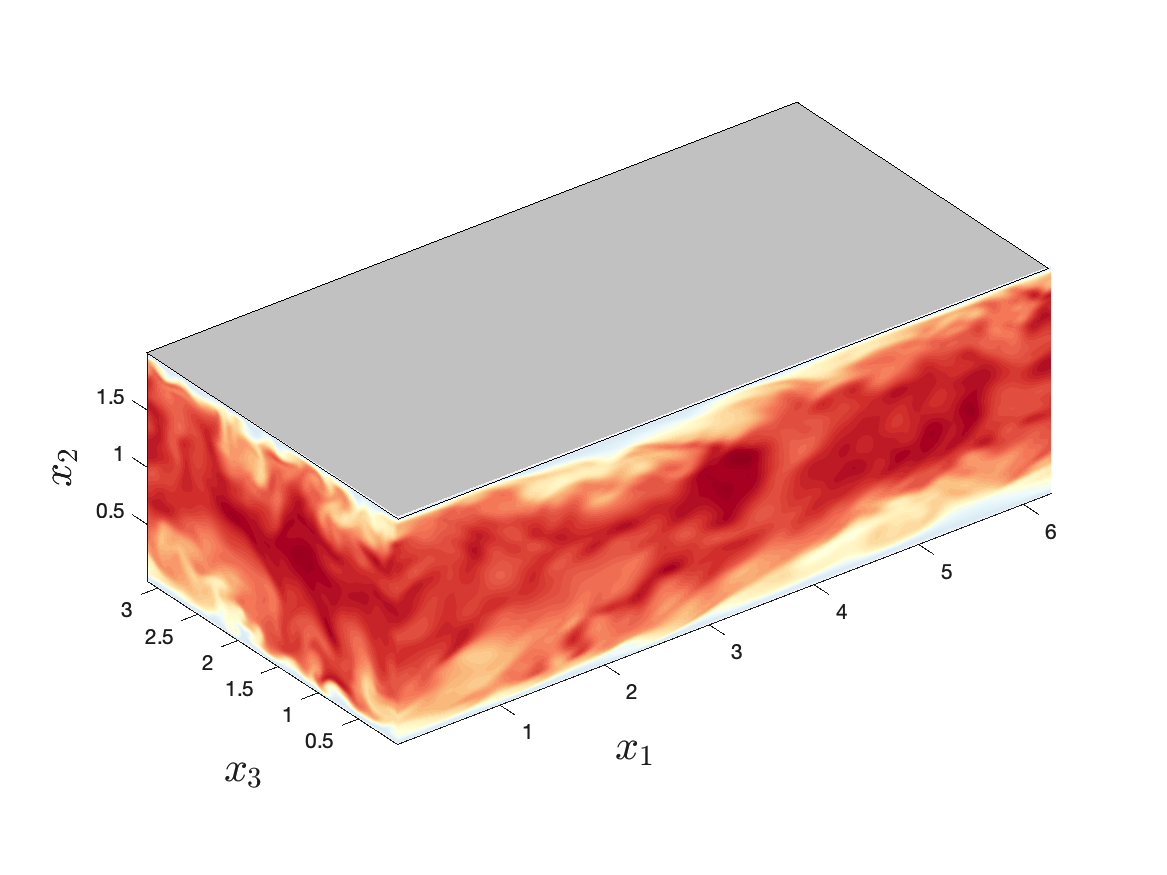}
\label{Fig:dataChannel}
}
\subfigure[Reconstructed HOSVD, $\overline{CR}=26.9$]{
\includegraphics[trim=0.5cm 0cm 1cm 0cm, clip=true, ,width=0.3\textwidth]{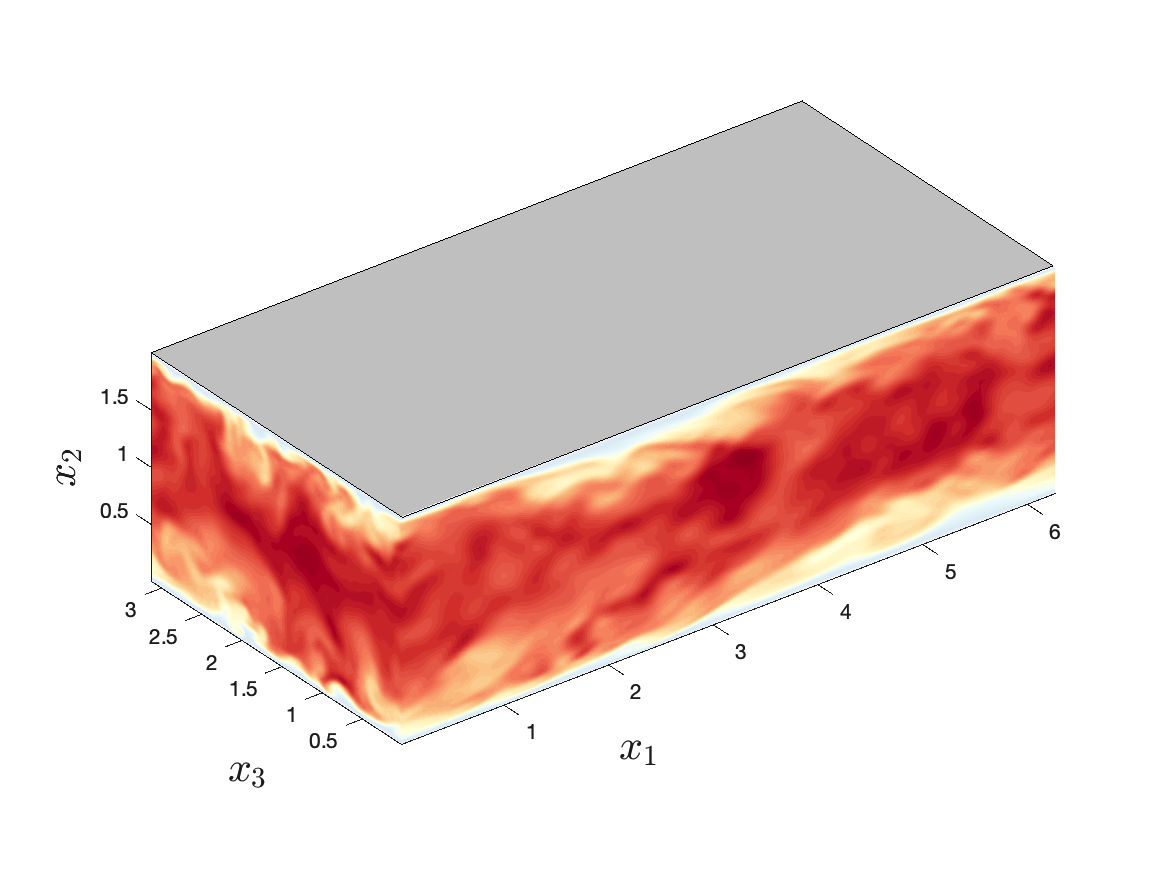}
\label{Fig:recdemo}
}
\subfigure[Reconstructed TDB, $\overline{CR}=26.9$]{
\includegraphics[trim=0.5cm 0cm 1cm 0cm, clip=true, ,width=0.3\textwidth]{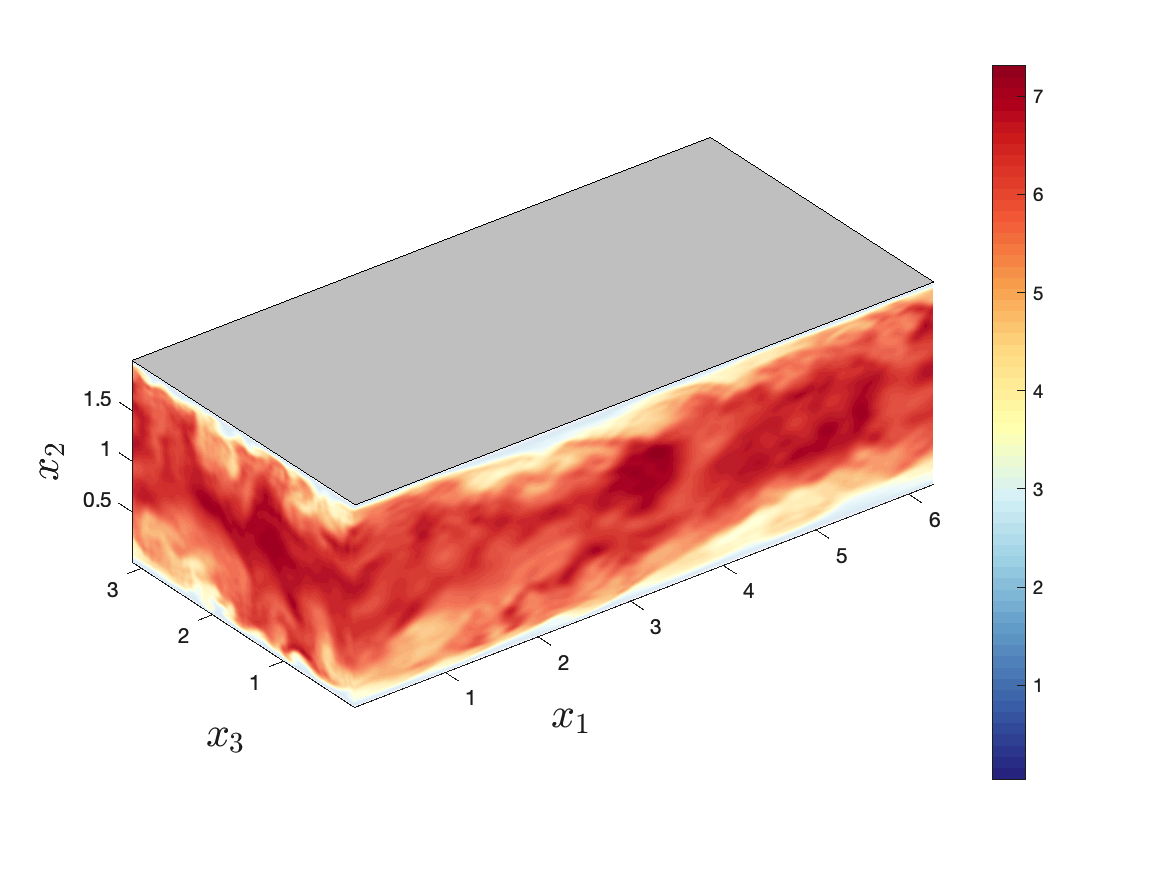}
\label{Fig:errordemo}
}
\caption{Turbulent channel flow:  (a) RMS, (b) mean profiles comparison between DNS and the same-rank HOSVD and TDB reductions. Comparison of  DNS and the same-rank HOSVD and TDB reductions: (c) DNS, (d) reconstructed TDB, and (e) reconstructed HOSVD.}
\label{Fig:ChannelMeanRMS}
\end{figure}
Figures \ref{Fig:RMS} and \ref{Fig:Mean} compare the $RMS$ and mean velocity at $t=1$ from TDB reconstructed calculations with Ref. \cite{vuorinen2016dnslab} and HOSVD reconstruction. In these figures, the TDB and HOSVD reconstructed $RMS$ and the mean velocity are in good agreement, however; due to unresolved modes the reconstructed data has discrepancy compared to the DNS results. We can observe the higher compression ratio is causing more discrepancy due to more unresolved modes. Figures \ref{Fig:dataChannel} to \ref{Fig:errordemo} compare the TDB and HOSVD reconstructed data with the streamed data for the case with high compression ratio at $t=1$ in 3D format. Based on these figures, we can conclude that both the HOSVD and TDB reconstruction are in good agreement with the original data.
% \begin{figure}[bp!]
% \centering
% \subfigure[DNS]{
% \includegraphics[trim=0.5cm 0cm 1cm 0cm, clip=true, width=0.3\textwidth]{ChannelF/3dChanneldat2.png}
% \label{Fig:dataChannel}
% }
% \subfigure[Reconstructed HOSVD, $CR=26.9$]{
% \includegraphics[trim=0.5cm 0cm 1cm 0cm, clip=true, ,width=0.3\textwidth]{ChannelF/3dChannelhosvd2.png}
% \label{Fig:recdemo}
% }
% \subfigure[Reconstructed TDB, $CR=26.9$]{
% \includegraphics[trim=0.5cm 0cm 1cm 0cm, clip=true, ,width=0.3\textwidth]{ChannelF/3dChannelrec2.png}
% \label{Fig:errordemo}
% }
% \caption{Turbulent Channel flow: Comparison of  DNS and the same-rank HOSVD and TDB reductions (a) DNS, (b) Reconstructed TDB, (c) Reconstructed HOSVD.}
% \label{Fig:ChannelData}
% \end{figure}
%

%%%%

%\section{Summary}\label{sec:summary}
%\input{Sections/Summary.tex}
\section{Summary}\label{sec:Disc}
We present an in situ compression method based on  TDB, in which the multidimensional streaming data are decomposed into a set of TDB and a time-dependent core tensor. We derived closed-form evolution equations for the  TDB and the core tensor.  The presented methodology is adaptive and maintains the error bellow the defined threshold $\varepsilon_{th}$ by adding/removing ranks. The computational cost of solving TDB computational complexity scales linearly with the data size, making it suitable for large-scale streaming datasets.\\

We perform this compression method on four cases:
\begin{enumerate}
\item Runge Function: Where we demonstrated the adaptive mode addition/removal. We also investigated the effect  of different time integration schemes on the compression error of TDB.
\item Incompressible Turbulent Reactive Flow: In this case, we compress the data by two TDB schemes (TDB-1 and TDB-2). TDB-1 has a higher compression ratio ($\overline{CR}=15.62$, which compresses $35GB$ to $2.2GB$) since it can decompose the physical space more than TDB-2 using one dimensional basis. TDB-2 scheme has a lower compression ratio ($\overline{CR}=6.6$, which compresses $35GB$ to $5.3GB$) because it uses a two dimensional basis for physical space. Since the physical space has a high dimensionality compared to composition space, the error growth rate for TDB-1 is higher and requires frequent reinitialization and mode adjustments. 
\item Incompressible Turbulent Reactive Flow with Random Diffusion Coefficient: This problem is the same as the second case with sixteen samples random diffusion coefficients. The TDB scheme in this problem is similar to the TDB-2 with an additional two dimensional basis for the random space. Since the TDB can exploit more correlation in the streamed data, the compression ratio is higher than the previous cases ($\overline{CR}=42.75$, which compresses $561.5GB$ to $13.1GB$).
\item Three-dimensional Turbulent Channel Flow:  In this problem, we compare two different compression ratios. We observed that the TDB with higher compression ratio has a slower growth rate compared to the TDB with lower compression ratio.  For both cases, the TDB reconstructed results are in good agreement with HOSVD. 
\end{enumerate}

 The current methodology only exploits multidimensional correlation in all dimensions except the temporal direction. For the future studies, we will extend this methodology to exploit temporal correlations. We will also pursue  building real-time reduced order models for diagnostic and predictive purposes.   
\section*{Acknowledgements}
This project was supported by an award by Air Force Office of Scientific Research (PM: Dr. Fariba Fahroo)  award FA9550-21-1-0247 and  support from NASA's Transformational Tools and Technologies Project (Cooperative Agreement No. 80NSSC18M0150). 
%\section{Appendix}\label{sec:App}
%\input{Sections/Appendix}
%\section{Appendix}\label{sec:App2}
%\input{Sections/Appendix2}
% \vspace{5mm}
%\printbibliography
\bibliographystyle{ieeetr}
%\bibliography{SHZA,HB}
\bibliography{main}
\newpage

\end{document}